\documentclass[pre,twocolumn,showpacs,superscriptaddress]{revtex4}
\usepackage{amsmath}
\usepackage{color}

\begin{document}

\title{A theory for the dynamics of dense systems of athermal self-propelled 
particles}

\author{Grzegorz Szamel}
\affiliation{Department of Chemistry, 
Colorado State University, Fort Collins, CO 80525}
\affiliation{Laboratoire Charles Coulomb, UMR 5221 CNRS,
Universit{\'e} Montpellier, Montpellier, France}

\date{\today}

\pacs{82.70.Dd, 64.70.pv, 64.70.Q-, 47.57.-s}

\begin{abstract}
We present a derivation of a recently proposed 
theory for the time dependence of density fluctuations in stationary
states  of strongly interacting, athermal, self-propelled particles. 
The derivation consists of two steps. 
First, we start from the equation of motion for the joint distribution of particles' 
positions and self-propulsions and we integrate out the self-propulsions. In this way we 
derive an approximate, many-particle evolution equation for the probability 
distribution of the particles' positions. 
Second, we use this evolution equation to describe the time dependence of steady-state 
density correlations. We derive a memory function representation of 
the density correlation function and then we use a factorization approximation to 
obtain an approximate expression for the memory function. In the final equation
of motion for the density correlation function the non-equilibrium character of the 
active system manifests itself through the presence of a new steady-state correlation
function that quantifies spatial correlations of the velocities of the particles. 
This correlation function enters into the frequency term, and thus it describes the
dependence of the short-time dynamics on the properties of
the self-propulsions. More importantly, the correlation function of particles' 
velocities enters into the vertex of the memory function and through the vertex it 
modifies the long-time glassy dynamics.
\end{abstract} 

\maketitle

\section{Introduction}\label{sec:intro}

There is a lot of interest in the static and dynamic 
properties of active matter systems 
\cite{Ramaswamyrev,Catesrev,Marchettirev}. These systems consist of particles 
that are self-propelled, and thus can move autonomously. The reason for this interest 
is twofold. First, there exists a great variety of biological and physical systems with
self-propelled constituents. In this work we consider only a subset of these systems  
consisting of assemblies of particles without aligning interactions. Biological
examples of such systems include some motile bacteria and physical systems are 
represented by synthetic Janus colloidal particles. The second reason for the 
interest in active matter systems is that they provide fascinating
examples of intrinsically non-equilibrium driven systems with very rich and unusual 
phenomenology. From the point of view of a theoretical analysis active systems
are a little easier to study than sheared systems, which used to be a workhorse of 
non-equilibrium statistical mechanics \cite{Fuchsrev}. This follows from the fact 
that in active matter systems the driving is applied locally and, on average, 
isotropically. In contrast, sheared systems are intrinsically anisotropic. 

Recently, it has been realized that dense active systems can exhibit glassy dynamics.
The analogy between the dynamics in dense cell layers and the dynamics in a fluid
approaching a glass transition was noted by Angelini \textit{et al.} 
\cite{Angelini}. This study inspired a simulational investigation 
of a two-dimensional dense active system with aligning interactions 
in which a jammed phase was identified \cite{HenkesFilyMarchetti}.
Glassy dynamics and the active glass transition was analyzed in a more general setting
by Berthier and Kurchan \cite{BerthierKurchan}. They studied
dynamic behavior of a simple model active system inspired by the spherical 
$p$-spin model. Their investigation showed that activity can shift the onset of glassy
behavior but it does not always destroy the glass transition, unlike for similar
models of sheared systems \cite{BBK}.
Berthier and Kurchan's theoretical analysis  
was followed by several computational studies of strongly interacting systems of
self-propelled particles without aligning interactions, 
which demonstrated glassy behavior of active matter. 
First, Ni \textit{et al.} \cite{Ni} simulated a system of
active Brownian hard spheres and found that, with increasing magnitude of the 
self-propulsion, the long-time dynamics speeds up whereas the local structure, monitored 
through the steady-state structure factor, becomes less pronounced. Next, Berthier 
\cite{Berthier} simulated a slightly different, fully athermal (\textit{i.e.} devoid
of thermal Brownian motion) system of active hard disks 
and found that, with increasing departure from equilibrium, 
at intermediate volume fractions the dynamics could be non-monotonic but, again,
the apparent glass transition is shifted towards higher volume fractions.
Wysocki \textit{et al.} \cite{Gompper} simulated a dense system of active Brownian 
particles with continuous interactions. 
More recently, Mandal \textit{et al.} \cite{Dasgupta} simulated the classic model
glassy system, the Kob-Andersen Lennard-Jones binary mixture \cite{Kob1994}, in
which one of the components was endowed with self-propulsion. In qualitative 
agreement with Ni \textit{et al.}, they found that, upon increasing the magnitude of the
self-propulsion, the long-time dynamics speeds up and that, beyond a critical strength
of the self-propulsion, the glass phase disappears.
Finally, Fily \textit{et al.} \cite{FilyHenkesMarchetti} investigated the 
phase diagram of active Brownian harmonic spheres as a function of
density, activity and noise, and identified a glassy phase in the high density, 
small self-propulsion speed regime. 

The simulational investigations of active glassy dynamics 
stimulated interest in theoretical
descriptions of the dynamics of dense systems of self-propelled particles. 
The main goal of these descriptions is to understand the effect of the 
non-equilibrium character of the active system on the glassy dynamics and the 
glass transition.  
We recall that, in spite of an extensive theoretical effort,  a microscopic and 
quantitatively correct description of glassy dynamics of thermal ``passive'' 
systems remains a challenge \cite{BerthierBiroli}. Thus, we should not expect to 
obtain a microscopic and quantitatively accurate theory for much more complicated 
active systems. However, we might be able to get some insight into the role of activity.
The precedent for this is the application of the  well-known, albeit far from 
perfect, theory of the glassy dynamics, the mode-coupling 
theory \cite{Goetzebook}, to the description of colloidal systems with short-range 
attractive interactions \cite{Dawson}. The theory, while in general not quantitatively 
correct, was able to account for the effect of attractions and to predict 
a re-entrant glass transition. Its qualitative predictions were subsequently 
confirmed by experiments \cite{Pham}. We note that proposed 
theoretical descriptions of the dynamics of dense active systems use 
approaches similar the mode-coupling theory. This is hardly surprising since the
mode-coupling theory provides one of very few microscopic descriptions 
of the glassy dynamics in thermal systems \cite{commentMCT}. 
In addition, the virtues and flaws of
this theory are quite well understood, which allows one to focus on phenomena that
the theory describes reasonably well.

The first microscopic theory for the glassy dynamics of dense systems 
consisting of spherically symmetric self-propelled particles with 
non-aligning interactions was put forward by Farage and 
Brader \cite{Farage}. This theory is a generalization of an earlier mode-coupling theory
for sheared glassy colloidal fluids proposed by Fuchs and Cates \cite{Fuchsrev}. 
Farage and Brader considered a system of self-propelled particles that were also 
subjected to thermal noise. They used the integration-through-transients approach 
introduced by Fuchs and Cates. In this approach one assumes that in the infinitely 
distant past the system was in an equilibrium state and then the drive, in this 
case the activity, was turned on. Using this procedure one can, in principle, calculate
both equal time and dynamic properties of active systems. However,
it is not clear how to use this approach to describe a fully athermal system
which does not have an equilibrium state without the drive. 

Recently, we proposed a different microscopic theory for dense athermal active systems 
\cite{activemct}. Like the theory of Farage and Brader, our theory relies upon
a factorization approximation, and thus it 
falls into a general category of mode-coupling-like approaches. 
There are, however, significant differences between our theory and that of 
Farage and Brader. First, our approach incorporates the influence of 
correlations between velocities of different particles on the dynamics of the
active system. This effect is neglected in the theory of Farage and Brader. Second, 
we use a projection operator defined through a steady-state distribution function. This
approach avoids using the integration-through-transients procedure and allows us to 
describe dynamics of fully athermal systems. 

In Ref. \cite{activemct} we presented an outline of the derivation of our theory. 
Here we present the details of the derivation and discuss all the approximations used. 
To make the present contribution self-contained we include some of the discussion
from the appendices of Ref. \cite{activemct}. In addition, we present a derivation 
of the theory for the time dependence of the tracer density fluctuations. 

The main object of our theory is the (collective) intermediate scattering function,
$F(q;t)$, which describes the time dependence of the collective density fluctuations,
\begin{equation}\label{Fqt0}
F(q;t) = \frac{1}{N}
\left<\sum_i e^{-i\mathbf{q}\cdot\mathbf{r}_i(t)}
\sum_j e^{i\mathbf{q}\cdot\mathbf{r}_j(0)} \right>.
\end{equation}
Here and in the following the brackets $\left< \dots \right>$ denote averaging
over a steady-state distribution of positions and self-propulsions. We will also
consider a closely related function, the self-intermediate scattering function,
which describes the time dependence of the tagged particle (tracer) 
density fluctuations,
$F_s(q;t)$,
\begin{equation}\label{Fsqt0}
F_s(q;t) = \frac{1}{N}
\left<\sum_i e^{-i\mathbf{q}\cdot\left(\mathbf{r}_i(t)-\mathbf{r}_i(0)\right)} \right>.
\end{equation}
The latter function allows us to calculate quantities that are usually monitored
in computer simulations, \textit{i.e} the relaxation time $\tau_\alpha$ defined 
through the relation \cite{commenttaus}
\begin{equation}
F_s(q;\tau_\alpha)=e^{-1},
\end{equation} 
and the (long-time) self-diffusion
coefficient $D$ defined through the hydrodynamic limit of the self-intermediate
scattering function, 
\begin{equation}
\lim_{t\to\infty,q\to 0, q^2t=\text{const.}} F_s(q;t)\sim \exp\left(-Dq^2 t\right).
\end{equation} 

The paper is organized as follows. In Section \ref{sec:model} we present and
motivate the model active system that we analyze in the reminder of the paper.
In Section \ref{sec:absc} we discuss the main assumption on which our approach relies, 
the absence of steady-state currents. In Section \ref{sec:effeom} we present the
derivation of the approximate equation of motion for the many-body distribution of
the positions of the active particles. This is followed by a brief discussion
(Section \ref{sec:corr})
of the short-time behavior of the scattering functions which illustrates 
the importance of the correlations between self-propulsions and positions. 
The derivations of the memory function representation and of the approximate
expression for the memory function are presented 
in sections \ref{sec:memf} and \ref{sec:mct}, respectively. In Section 
\ref{sec:glassydyn} we discuss general conclusions that can be drawn from the 
final equation for the time dependence of the density correlations 
and in Section \ref{sec:disc} we 
summarize our findings and outline directions of future work. In the first appendix,
Appendix \ref{ap:alter}, 
we discuss an alternative equation of motion for the many-body distribution of
the particles' positions. In the second appendix,
Appendix \ref{ap:self}, we present the derivation of the theory for the time dependence
of the tagged particle (tracer) density fluctuations.

\section{Model active system}\label{sec:model}

We consider a system of $N$ interacting, self-propelled particles in a volume $V$. 
The average density is $\rho=N/V$. The particles interact via a spherically
symmetric potential $V(r)$. They move in a viscous medium that is characterized
by the friction coefficient of a single particle, which we denote by $\xi_0$. 
We assume that the friction felt by a particle
is independent of the particle density and configuration, and thus we neglect 
hydrodynamic interactions \cite{Dhont}. 
Each particle moves under the combined influence of the interparticle force
derived from the potential $V(r)$ and a self-propulsion force \cite{commentSP}.

In the active Brownian particles model in Refs.~\cite{tenHagen,FilyMarchetti} 
it is assumed that the 
magnitude of the self-propulsion force is constant whereas its direction changes 
\textit{via} rotational Brownian motion. If needed, to distinguish this model from other 
models of active particles we will refer to it as the rotational diffusion active
Brownian particles model. 

We use a slightly different
model of active particles, introduced in Ref. \cite{toy} and then analyzed in
Ref. \cite{activemct}. An essentially identical model was independently introduced and
studied by Maggi \textit{et al.} \cite{Maggi}. 

Specifically, we assume that the self-propulsion force does not have 
a constant magnitude and it evolves in time according to the Ornstein-Uhlenbeck process. 
We note that the rotational diffusion active Brownian particles model and our
model can be considered to be limiting cases of a more general model in which 
the direction of the self-propulsion force randomly rotates and its magnitude
fluctuates. Most likely,
neither limiting case perfectly describes active particles studied in 
experiments. The advantage of our model is that the equations of motion for the
self-propulsion are linear, which makes some theoretical considerations somewhat
easier. In particular, we showed that the complete 
steady-state distribution can be found for a single particle in a harmonic potential 
for our model \cite{toy}.
We note that our model can be considered to be a continuous time 
version of the model used by Berthier \cite{Berthier}. Finally, we mention that a
mapping was derived between the rotational diffusion active 
Brownian particles model and our model \cite{FilyMarchetti,FarageKB}. This mapping 
preserves the self-propulsion autocorrelation function. However, higher moments of 
the self-propulsion are different.  

We assume that the self-propelled particles are large enough so that any Brownian motion 
due to the thermal fluctuations of the viscous medium (solvent) can be neglected. 
Thus, the system is purely athermal and the particles move under the sole influence
of the interparticle interactions and self-propulsion. 

The above qualitative discussion translates into the following equations of 
motion for the positions and self-propulsions, 
\begin{eqnarray}\label{eompos}
\dot{\mathbf{r}}_i &=& \xi_0^{-1}\left[\mathbf{f}_i  + \mathbf{F}_i
\right], \\ \label{eomsp}
\dot{\mathbf{f}}_i &=& -\tau_p^{-1} \mathbf{f}_i + \boldsymbol{\eta}_i.
\end{eqnarray}
In Eq.~(\ref{eompos}), $\mathbf{r}_i$ is the position of particle $i$,
$\xi_0$ the friction coefficient of an isolated particle,
$\mathbf{f}_i$ is the self-propulsion acting on particle $i$ and
$\mathbf{F}_i$ is the force acting on particle $i$ originating 
from the interactions, 
\begin{equation}
\mathbf{F}_i = -\sum_{j\neq i} \boldsymbol{\nabla}_i V(r_{ij}).
\end{equation}  
In Eq. (\ref{eomsp}), 
$\tau_p$ is the persistence time of the self-propulsion and
$\boldsymbol{\eta}_i$ is an internal Gaussian noise with zero mean and 
variance 
\begin{equation}
\left<\boldsymbol{\eta}_i(t) \boldsymbol{\eta}_j(t')
\right>_{\text{noise}} = 
2 D_f\boldsymbol{I}\delta_{ij}\delta(t-t'),
\end{equation} 
where $\left< ... \right>_{\text{noise}}$ 
denotes averaging over the noise distribution, $D_f$ is the noise strength
and $\boldsymbol{I}$ is the unit tensor.
Without interactions, particles evolving according to
Eqs. (\ref{eompos}-\ref{eomsp}) perform a persistent random 
walk with the mean-square displacement \cite{VanKampen}
\begin{equation}\label{msdfree}
\left<\left(\mathbf{r}_i(t)-\mathbf{r}_i(0)\right)^2\right> = 
6 \frac{D_f \tau_p^2}{\xi_0^2}\left(t+\tau_p(e^{-t/\tau_p}-1)\right).
\end{equation}
According to Eq. (\ref{msdfree}), the short-time motion is ballistic,
\begin{equation}\label{msdfreess}
\left<\left(\mathbf{r}_i(t)-\mathbf{r}_i(0)\right)^2\right> \approx
3 \frac{D_f \tau_p}{\xi_0^2} t^2 \;\;\;\;\; t\ll \tau_p
\end{equation}
and the long-time motion, $t\gg \tau_p$, is diffusive with diffusion coefficient $D_0$,
\begin{equation}\label{difffree}
D_0 = D_f\tau_p^2/\xi_0^2.
\end{equation}
Comparing expression (\ref{difffree}) with the well-known formula for the 
diffusion coefficient of a Brownian particle moving in a viscous medium with 
friction constant $\xi_0$, $D_\text{Brownian} = T/\xi_0$ (we use units such that 
the Boltzmann constant $k_B=1$), we can 
define the \textit{single-particle} effective temperature \cite{toy},
\begin{equation}\label{Teff}
T_{\text{eff}}= D_0 \xi_0  = D_f \tau_p^2/\xi_0. 
\end{equation}
The single particle effective temperature could be used as one of the independent
control parameters, together with the number density $\rho$ and the persistence time 
$\tau_p$. 

Eqs. (\ref{eompos}-\ref{eomsp}) are a convenient starting point for computer 
simulations. In theoretical considerations it is more useful to describe the system 
using probability distributions and their associated evolution equations. 
The most fundamental description of our system, 
equivalent to equations of motion (\ref{eompos}-\ref{eomsp}), uses the
joint $N$-particle probability distribution of positions and self-propulsions,
$P_N(\mathbf{r}_1,\mathbf{f}_1,...,\mathbf{r}_N,\mathbf{f}_N;t)$.  This 
distribution evolves in time with evolution operator $\Omega$ 
\begin{equation}\label{eom}
\partial_t P_N(\mathbf{r}_1,\mathbf{f}_1,...,\mathbf{r}_N,\mathbf{f}_N;t) =
\Omega P_N(\mathbf{r}_1,\mathbf{f}_1,...,\mathbf{r}_N,\mathbf{f}_N;t),
\end{equation} 
which can be derived \cite{VanKampen} from equations of motion 
(\ref{eompos}-\ref{eomsp}),
\begin{equation}\label{Omega1}
\Omega = -\xi_0^{-1} \sum_{i} \boldsymbol{\nabla}_i\cdot 
\left(\mathbf{f}_i + \mathbf{F}_i \right) 
+ \sum_{i} \frac{\partial}{\partial \mathbf{f}_i}\cdot
\left( \tau_p^{-1} \mathbf{f}_i  + D_f \frac{\partial}{\partial 
\mathbf{f}_i}\right).
\end{equation}

We recall that for non-interacting active particles Eqs. (\ref{eom}-\ref{Omega1})
are formally equivalent to the 
Fokker-Planck equation that describes the motion of non-interacting 
Brownian particles on a time scale on which their velocity
relaxation can be observed \cite{VanKampen}. Thus, a system of non-interacting 
active particles is formally equivalent to particles moving under the influence of  
thermal solvent fluctuations. This equivalence is absent for a system of
\textit{interacting} active particles. In particular, it should be emphasized that 
according to Eqs. (\ref{eompos}-\ref{eomsp}) or, equivalently, Eqs. 
(\ref{eom}-\ref{Omega1}), self-propulsions evolves autonomously, independently of the 
configuration of the particles.

We will assume that our system can achieve a stationary state. In other words, there
exists a steady-state probability distribution 
$P_N^{\text{ss}}(\mathbf{r}_1,\mathbf{f}_1,...,\mathbf{r}_N,\mathbf{f}_N)$ such
that
\begin{equation}\label{ss}
\Omega P_N^{\text{ss}}(\mathbf{r}_1,\mathbf{f}_1,...,\mathbf{r}_N,\mathbf{f}_N) = 0.
\end{equation}

We emphasize that, in general, the joint steady-state distribution of
positions and self-propulsions does not
factorize into a product of steady-state distributions of particle positions 
and self-propulsions,
\begin{equation}
P_N^{\text{ss}}(\mathbf{r}_1,\mathbf{f}_1,...,\mathbf{r}_N,\mathbf{f}_N)
\neq
P_N^{\text{ss}}(\mathbf{r}_1,...,\mathbf{r}_N)
P_N^{\text{ss}}(\mathbf{f}_1,...,\mathbf{f}_N),
\end{equation}
where $P_N^{\text{ss}}(\mathbf{r}_1,...,\mathbf{r}_N)$
and $P_N^{\text{ss}}(\mathbf{f}_1,...,\mathbf{f}_N)$ are the steady-state 
distributions of positions and self-propulsions,
\begin{eqnarray}
P_N^{\text{ss}}(\mathbf{r}_1,...,\mathbf{r}_N) \!\!\! &=& \!\!\! 
\int  d\mathbf{f}_1 ... d\mathbf{f}_N 
P_N^{\text{ss}}(\mathbf{r}_1,\mathbf{f}_1,...,\mathbf{r}_N,\mathbf{f}_N),
\\
P_N^{\text{ss}}(\mathbf{f}_1,...,\mathbf{f}_N) \!\!\! &=& \!\!\! 
\int  d\mathbf{r}_1 ... d\mathbf{r}_N 
P_N^{\text{ss}}(\mathbf{r}_1,\mathbf{f}_1,...,\mathbf{r}_N,\mathbf{f}_N).
\end{eqnarray}
In general, neither the joint steady-state distribution 
$P_N^{\text{ss}}(\mathbf{r}_1,\mathbf{f}_1,...,\mathbf{r}_N,\mathbf{f}_N)$
nor the steady-state distributions of positions
$P_N^{\text{ss}}(\mathbf{r}_1,...,\mathbf{r}_N)$ are known exactly (for approximate
theories for the latter distribution 
see Refs.~\cite{Maggi,FarageKB}). However, the steady-state 
distribution of self-propulsions has a simple form,
\begin{equation}
P_N^{\text{ss}}(\mathbf{f}_1,...,\mathbf{f}_N) \propto
\exp\left(-\frac{\sum_i\mathbf{f}^2_i}{2 D_f \tau_p}\right).
\end{equation}

Evolution operator (\ref{Omega1}) allows us to rewrite the definitions of
the intermediate scattering functions (\ref{Fqt0}-\ref{Fsqt0}),
\begin{equation}\label{Fqt}
F(q;t) = \frac{1}{N}
\left<n(\mathbf{q}) \exp\left(\Omega t\right) n(-\mathbf{q})\right>,
\end{equation}
\begin{equation}\label{Fsqt}
F_s(q;t) = 
\left<n_s(\mathbf{q}) \exp\left(\Omega t\right) n_s(-\mathbf{q})\right>.
\end{equation}
In Eq. (\ref{Fqt}) $n(\mathbf{q})$  
is the Fourier transform of the microscopic density,
\begin{equation}\label{n1def}
n(\mathbf{q}) = \sum_l e^{-i\mathbf{q}\cdot\mathbf{r}_l},
\end{equation}
and in Eq. (\ref{Fsqt}) $n_s(\mathbf{q})$ is the Fourier transform of the microscopic 
tagged particle (tracer) density,
\begin{equation}\label{nsdef}
n_s(\mathbf{q}) = e^{-i\mathbf{q}\cdot\mathbf{r}_1}.
\end{equation}
Note that we chose the tagged particle to be the particle number one; this choice
is made arbitrarily and for convenience only. 
In Eqs. (\ref{Fqt}-\ref{Fsqt}) $\Omega$ is the evolution operator (\ref{Omega1}), 
and brackets $\left<\dots\right>$ denote
averaging over the joint steady-state distribution 
of positions and self-propulsions (\ref{ss}). We emphasize that in 
Eqs. (\ref{Fqt}-\ref{Fsqt}) and in all similar formulas the steady-state distribution 
stands to the right of the quantity being averaged, and all operators act on it too. 

\section{The main assumption: absence of currents}\label{sec:absc}

The fundamental difficulty posed by driven systems originates from the absence of 
detailed balance. The lack of detailed balance allows for the existence of non-trivial 
currents. It has been argued that at a mesoscopic level, 
in some systems without aligning interactions, 
\textit{i.e.} after some coarse-graining (\textit{i.e.} above a certain
length and time scale), there are no
currents and the active system becomes equivalent to a passive system 
\cite{CatesTailleur}.

Our main assumption is that in our model system, 
in the steady state the currents vanish after integrating out
the self-propulsions. This assumption, together with
additional coarse-graining over time, will allow us to approximate our system by a 
passive system with detailed balance. 

A similar assumption was implicitly used by Farage and Brader. However, as 
discussed in the next section, their theory and ours use different passive systems to
approximate an active system. 
In particular, we retain correlations between (overdamped) velocities of 
different particles. 

To make our assumption explicit we first rewrite the equation of motion 
for the joint probability distribution
of positions and self-propulsions, Eq. (\ref{eom}), in the form of a continuity
equation,
\begin{eqnarray}\label{cont}
\lefteqn{ 
\partial_t P_N(\mathbf{r}_1,\mathbf{f}_1,...,\mathbf{r}_N,\mathbf{f}_N;t) = }
\\ \nonumber && 
-\sum_{i} \boldsymbol{\nabla}_i\cdot 
\mathbf{j}_i(\mathbf{r}_1,\mathbf{f}_1,...,\mathbf{r}_N,\mathbf{f}_N;t)
\\ \nonumber && 
-\sum_{i} \frac{\partial}{\partial \mathbf{f}_i}\cdot 
\mathbf{j}_i^{\mathbf{f}}(\mathbf{r}_1,\mathbf{f}_1,...,\mathbf{r}_N,\mathbf{f}_N;t),
\end{eqnarray}
where current densities are defined as
\begin{eqnarray}\label{currentpos0}
\lefteqn{
\mathbf{j}_i(\mathbf{r}_1,\mathbf{f}_1,...,\mathbf{r}_N,\mathbf{f}_N;t) = }
\\ \nonumber && 
\xi_0^{-1} 
\left(\mathbf{F}_i + \mathbf{f}_i \right)
P_N(\mathbf{r}_1,\mathbf{f}_1,...,\mathbf{r}_N,\mathbf{f}_N;t),
\end{eqnarray}
\begin{eqnarray}\label{currentsp0}
\lefteqn{
\mathbf{j}_i^{\mathbf{f}}(\mathbf{r}_1,\mathbf{f}_1,...,\mathbf{r}_N,\mathbf{f}_N;t) = }
\\ \nonumber && 
-\left( \tau_p^{-1} \mathbf{f}_i  + D_f \frac{\partial}{\partial 
\mathbf{f}_i}\right)
P_N(\mathbf{r}_1,\mathbf{f}_1,...,\mathbf{r}_N,\mathbf{f}_N;t).
\end{eqnarray}

Current densities (\ref{currentpos0}-\ref{currentsp0}) are microscopic
quantities, which may be non-zero in a system without detailed balance. 
Our main assumption is that in the steady state, 
the current density in the position space, integrated over self-propulsions, 
vanishes,
\begin{eqnarray}\label{curvan}
&& \mathbf{j}_i^{\text{ss}}(\mathbf{r}_1,...,\mathbf{r}_N;t) = 
\xi_0^{-1}\int  d\mathbf{f}_1 ... d\mathbf{f}_N 
\mathbf{j}_i^{\text{ss}}(\mathbf{r}_1,\mathbf{f}_1,...,\mathbf{r}_N,\mathbf{f}_N;t)
\nonumber \\ && \equiv 
\xi_0^{-1}\int  d\mathbf{f}_1 ... d\mathbf{f}_N
\left[\mathbf{F}_i + \mathbf{f}_i \right]
P_N^{\text{ss}}(\mathbf{r}_1,\mathbf{f}_1,...,\mathbf{r}_N,\mathbf{f}_N)
\nonumber \\ && = 0.
\end{eqnarray}

Assumption (\ref{curvan}) implies that the local steady-state average of 
the self-propulsion is equal to the negative of the force,
\begin{equation}\label{avsp}
\left<\mathbf{f}_i\right>_{\text{lss}} = - \mathbf{F}_i,
\end{equation}
where
the local steady-state average is defined as
\begin{eqnarray}
\lefteqn{ \left< \dots \right>_{\text{lss}} = }
\\ \nonumber && 
\frac{1}{P_N^{\text{ss}}(\mathbf{r}_1,...,\mathbf{r}_N)}
\int  d\mathbf{f}_1 ... d\mathbf{f}_N
\dots
P_N^{\text{ss}}(\mathbf{r}_1,\mathbf{f}_1,...,\mathbf{r}_N,\mathbf{f}_N).
\end{eqnarray}
Eq. (\ref{avsp}) could be interpreted as stating a balance of the self-propulsion
of particle $i$ and the total potential force acting on this particle, for a 
given configuration of the system.  

We are  not aware of any study that specifically focused on the existence
of non-trivial steady-state currents in high density active systems without aligning 
interactions \cite{commentcurrents}. 
We note that the assumption (\ref{curvan}) is made at the level of $N$-particle 
quantities. Thus, its direct simulational verification seems rather difficult.
However, it might be possible to define and measure reduced (few-particle) current 
densities. Alternatively, it might be possible to analyze theoretically and then
verify computationally the consequences of the presence/absence of currents. 
For example, one could try to derive a theory for an effective temperature 
of the active system and then to compare it with a recent simulational investigation
\cite{LevisBerthier}. Finally, one could try to develop an approximate expression
for the $N$-particle joint steady-state distribution of positions and self-propulsions
and investigate the validity of Eq. (\ref{curvan}) directly. Preliminary results from 
such a project \cite{privatecomm} suggest that, in general, steady-state currents exist 
and thus Eq. (\ref{curvan}) constitutes an approximation. 
We will return to the consequence of the assumption made in 
Eq. (\ref{curvan}) in future work.

\section{Effective equation of motion for the distribution
of particle positions}\label{sec:effeom}

As emphasized in Sec. \ref{sec:model}, in our model active system the evolution
of the self-propulsions does not depend on the configuration and density of the
particles. On the other hand, for any interparticle interactions with a strongly
repulsive short-range part of the interaction potential, in a dense system 
the positions of the particles change slowly. Intuitively, upon increasing the
strength of the interactions in a dense active system, one expects qualitatively
similar slowing down as in a passive system. 
By and large, this expectation has been borne out by computer simulations 
\cite{Ni,Berthier,Dasgupta,activemct} albeit with some caveats.

Thus, it seems reasonable to assume that in strongly interacting systems of
active particles without aligning interactions 
the self-propulsions relax faster than the positions of the particles.
This separation of times scales suggests that it should be possible to derive
an approximate equation of motion for the probability of particles' positions only.
We expect that such an equation should become progressively more accurate in the 
strong interaction limit. 

We should recall that our goal is to describe the dynamics of density 
fluctuations in the steady state of our model active system. Thus, we only need an 
approximate equation of motion for the probability of particles' positions 
in the vicinity of the steady state. For this reason, in our considerations we
introduce the steady-state distribution and, effectively, we use the gradient
of the logarithm of this distribution with respect to the position of a given particle 
as a (normalized) effective force acting on this particle (see Eq. (\ref{Feff})). 
This construction
is different from the approach of Farage and Brader \cite{Farage} who concentrate
on transient density fluctuations. 

To facilitate the derivation of the equation of motion for the probability of 
particles' positions we introduce a projection operator that acts on
an $N$-particle probability distribution of self-propulsions and positions and
projects it on a local steady-state distribution, \textit{i.e.} on a distribution 
in which self-propulsions have a steady-state distribution for a given sample 
of positions,
\begin{eqnarray}
\lefteqn{ \mathcal{P}_{\text{lss}} 
P_N(\mathbf{r}_1,\mathbf{f}_1,...,\mathbf{r}_N,\mathbf{f}_N;t)}
\nonumber \\ 
&=& 
\frac{P_N^{\text{ss}}(\mathbf{r}_1,\mathbf{f}_1,...,\mathbf{r}_N,\mathbf{f}_N)}
{P_N^{\text{ss}}(\mathbf{r}_1,...,\mathbf{r}_N)}
\int  d\mathbf{f}_1 ... d\mathbf{f}_N
P_N(\mathbf{r}_1,\mathbf{f}_1,...,\mathbf{r}_N,\mathbf{f}_N;t)
\nonumber \\ 
&=& \frac{P_N^{\text{ss}}(\mathbf{r}_1,\mathbf{f}_1,...,\mathbf{r}_N,\mathbf{f}_N)}
{P_N^{\text{ss}}(\mathbf{r}_1,...,\mathbf{r}_N)} 
P_N(\mathbf{r}_1,...,\mathbf{r}_N;t).
\end{eqnarray}
We note that by integrating $ \mathcal{P}_{\text{lss}} 
P_N(\mathbf{r}_1,\mathbf{f}_1,...,\mathbf{r}_N,\mathbf{f}_N;t)$ over self-propulsions
we get the probability distribution of particles' positions, 
$P_N(\mathbf{r}_1,...,\mathbf{r}_N;t)$.

Next, we define the orthogonal projection,
\begin{eqnarray}
\mathcal{Q}_{\text{lss}} = \mathcal{I} - \mathcal{P}_{\text{lss}},
\end{eqnarray}
and write down equations of motion for 
$\mathcal{P}_{\text{lss}} 
P_N(\mathbf{r}_1,\mathbf{f}_1,...,\mathbf{r}_N,\mathbf{f}_N;t)$ and
$\mathcal{Q}_{\text{lss}} 
P_N(\mathbf{r}_1,\mathbf{f}_1,...,\mathbf{r}_N,\mathbf{f}_N;t)$,
\begin{eqnarray}\label{PPeom}
\lefteqn{
\partial_t \mathcal{P}_{\text{lss}} 
P_N(\mathbf{r}_1,\mathbf{f}_1,...,\mathbf{r}_N,\mathbf{f}_N;t)=}
\nonumber \\ && 
\mathcal{P}_{\text{lss}} \Omega \mathcal{P}_{\text{lss}} 
P_N(\mathbf{r}_1,\mathbf{f}_1,...,\mathbf{r}_N,\mathbf{f}_N;t) 
\nonumber \\ && + 
\mathcal{P}_{\text{lss}} \Omega \mathcal{Q}_{\text{lss}} 
P_N(\mathbf{r}_1,\mathbf{f}_1,...,\mathbf{r}_N,\mathbf{f}_N;t),
\end{eqnarray}
\begin{eqnarray}\label{QPeom}
\lefteqn{
\partial_t \mathcal{Q}_{\text{lss}} 
P_N(\mathbf{r}_1,\mathbf{f}_1,...,\mathbf{r}_N,\mathbf{f}_N;t) = }
\nonumber \\ && 
\mathcal{Q}_{\text{lss}} \Omega \mathcal{P}_{\text{lss}} 
P_N(\mathbf{r}_1,\mathbf{f}_1,...,\mathbf{r}_N,\mathbf{f}_N;t) 
\nonumber \\ && 
+ \mathcal{Q}_{\text{lss}} \Omega \mathcal{Q}_{\text{lss}} 
P_N(\mathbf{r}_1,\mathbf{f}_1,...,\mathbf{r}_N,\mathbf{f}_N;t).
\end{eqnarray}

Since our goal is to calculate the intermediate scattering functions,
Eqs. (\ref{Fqt}-\ref{Fsqt}), which are functions of positions only, we can assume that 
\begin{equation}
\mathcal{Q}_{\text{lss}}
P_N(\mathbf{r}_1,\mathbf{f}_1,...,\mathbf{r}_N,\mathbf{f}_N;t=0)=0.
\end{equation} 
Then we  can solve Eqs. (\ref{PPeom}-\ref{QPeom}) for 
the Laplace transform, $\mathcal{LT}$, of 
$\partial_t \mathcal{P}_{\text{lss}} 
P_N(\mathbf{r}_1,\mathbf{f}_1,...,\mathbf{r}_N,\mathbf{f}_N;t)$ and we obtain
\begin{widetext}
\begin{eqnarray}\label{proj1}
\mathcal{LT}\left[\partial_t \mathcal{P}_{\text{lss}} 
P_N(\mathbf{r}_1,\mathbf{f}_1,...,\mathbf{r}_N,\mathbf{f}_N;t)\right](z) = 
\left[  \mathcal{P}_{\text{lss}} \Omega \mathcal{P}_{\text{lss}} 
+ \mathcal{P}_{\text{lss}} \Omega \mathcal{Q}_{\text{lss}}
\frac{1}{z-\mathcal{Q}_{\text{lss}} \Omega \mathcal{Q}_{\text{lss}}} 
\mathcal{Q}_{\text{lss}}\Omega\mathcal{P}_{\text{lss}} \right]
\mathcal{P}_{\text{lss}} P_N(\mathbf{r}_1,\mathbf{f}_1,...,\mathbf{r}_N,\mathbf{f}_N;z).
\end{eqnarray}
The first term inside the brackets on right-hand-side of Eq. (\ref{proj1}) reads
\begin{eqnarray}\label{proj1firstterm}
\mathcal{P}_{\text{lss}} \Omega \mathcal{P}_{\text{lss}} 
P_N(\mathbf{r}_1,\mathbf{f}_1,...,\mathbf{r}_N,\mathbf{f}_N;z) &=&
\frac{P_N^{\text{ss}}(\mathbf{r}_1,\mathbf{f}_1,...,\mathbf{r}_N,\mathbf{f}_N)}
{P_N^{\text{ss}}(\mathbf{r}_1,...,\mathbf{r}_N)}
\xi_0^{-1}\int  d\mathbf{f}_1 ... d\mathbf{f}_N
\left[\mathbf{F}_i + \mathbf{f}_i \right]
P_N^{\text{ss}}(\mathbf{r}_1,\mathbf{f}_1,...,\mathbf{r}_N,\mathbf{f}_N)\cdot
\nonumber \\ && \times
\boldsymbol{\nabla}_i \frac{P_N(\mathbf{r}_1,...,\mathbf{r}_N;z)}
{P_N^{\text{ss}}(\mathbf{r}_1,...,\mathbf{r}_N)}
\end{eqnarray}
We see that if current densities vanish in the steady state, 
Eq. (\ref{curvan}), this term vanishes. 
Furthermore, one can show that
\begin{eqnarray}\label{right}
\mathcal{Q}_{\text{lss}}\Omega \mathcal{P}_{\text{lss}} P_N(z) =
- \xi_0^{-1} \sum_i 
\left(\mathbf{f}_i - \left<\mathbf{f}_i\right>_{\text{lss}}\right)
P_N^{\text{ss}}(\mathbf{r}_1,\mathbf{f}_1,...,\mathbf{r}_N,\mathbf{f}_N)
\cdot
\left[\boldsymbol{\nabla}_i \frac{P_N(\mathbf{r}_1,...,\mathbf{r}_N;z)}
{P_N^{\text{ss}}(\mathbf{r}_1,...,\mathbf{r}_N)}\right]
\end{eqnarray}
and
\begin{eqnarray}\label{left}
\mathcal{P}_{\text{lss}} \Omega \mathcal{Q}_{\text{lss}} ...  =  
- \frac{P_N^{\text{ss}}(\mathbf{r}_1,\mathbf{f}_1,...,\mathbf{r}_N,\mathbf{f}_N)}
{P_N^{\text{ss}}(\mathbf{r}_1,...,\mathbf{r}_N)} 
\xi_0^{-1} \sum_i \boldsymbol{\nabla}_i \cdot
\int  d\mathbf{f}_1 ... d\mathbf{f}_N
\left(\mathbf{f}_i - \left<\mathbf{f}_i\right>_{\text{lss}}\right) ...
\end{eqnarray}
\end{widetext}
So far, we have not made any approximations. To proceed, we 
will need to deal with projected evolution operator 
$\mathcal{Q}_{\text{lss}}\Omega\mathcal{Q}_{\text{lss}}$ in Eq. (\ref{proj1}). 
This operator 
describes evolution in the space orthogonal to the local steady-state space. The 
simplest possible approximation is to
assume that this evolution is entirely due to the free relaxation of
the self-propulsions. In this case 
$\mathcal{Q}_{\text{lss}}\Omega\mathcal{Q}_{\text{lss}}$ is approximated as 
follows
\begin{equation}\label{appQOQ}
\mathcal{Q}_{\text{lss}}\Omega\mathcal{Q}_{\text{lss}} \approx
\sum_{i=1}^{N} \frac{\partial}{\partial \mathbf{f}_i}
\left( \tau_p^{-1} \mathbf{f}_i  + D_f \frac{\partial}{\partial \mathbf{f}_i}\right).
\end{equation}
We note that approximation (\ref{appQOQ}) is equivalent to
assuming that the time-scale of the self-propulsion's relaxation is
the same in non-interacting and interacting systems. This can
be approximately valid for our model active system but, in particular, it 
would be an unreasonable approximation in the presence of aligning
interactions. We recall that in our system the evolution of the self-propulsions
is independent of the positions of the particles. Thus, the approximation (\ref{appQOQ}) 
neglects the influence of the correlations between self-propulsions and 
positions on the evolution of the self-propulsions. 
Combining this approximation with Eqs. (\ref{right}-\ref{left}) we get the following
approximate equality
\begin{widetext}
\begin{eqnarray} 
&&
\mathcal{P}_{\text{lss}} \Omega \mathcal{Q}_{\text{lss}}
\left(z-\mathcal{Q}_{\text{lss}}\Omega\mathcal{Q}_{\text{lss}}\right)^{-1} 
\mathcal{Q}_{\text{lss}}\Omega \mathcal{P}_{\text{lss}} 
P_N(\mathbf{r}_1,\mathbf{f}_1,...,\mathbf{r}_N,\mathbf{f}_N;z)
\approx
\frac{P_N^{\text{ss}}(\mathbf{r}_1,\mathbf{f}_1,...,\mathbf{r}_N,\mathbf{f}_N)}
{P_N^{\text{ss}}(\mathbf{r}_1,...,\mathbf{r}_N)} 
\xi_0^{-2} \sum_i \boldsymbol{\nabla}_i \cdot
\int  d\mathbf{f}_1 ... d\mathbf{f}_N
\left(\mathbf{f}_i - \left<\mathbf{f}_i\right>_{\text{lss}}\right)
\nonumber \\ && 
\left[z - \sum_{j=1}^{N} \frac{\partial}{\partial \mathbf{f}_j}
\left( \tau_p^{-1} \mathbf{f}_j  + D_f \frac{\partial}{\partial \mathbf{f}_j}\right)
\right]^{-1}
\sum_l
\left(\mathbf{f}_l - \left<\mathbf{f}_l\right>_{\text{lss}}\right)
P_N^{\text{ss}}(\mathbf{r}_1,\mathbf{f}_1,...,\mathbf{r}_N,\mathbf{f}_N)
\cdot
\left[\boldsymbol{\nabla}_l \frac{P_N(\mathbf{r}_1,...,\mathbf{r}_N;z)}
{P_N^{\text{ss}}(\mathbf{r}_1,...,\mathbf{r}_N)}\right].
\end{eqnarray}
Now, we expand $\left[z - \sum_{i=1}^{N} \frac{\partial}{\partial \mathbf{f}_i}
\left( \tau_p^{-1} \mathbf{f}_i  + D_f \frac{\partial}{\partial \mathbf{f}_i}\right)
\right]^{-1}$ and integrate by parts.  Finally, we integrate both sides of
the resulting equation over self-propulsions and get the following expression
for the Laplace transform of $\partial_t P_N(\mathbf{r}_1,...,\mathbf{r}_N;t)$
\begin{eqnarray}\label{proj2}
\mathcal{LT}
\left[\partial_t P_N(\mathbf{r}_1,...,\mathbf{r}_N;t)\right](z) =
\xi_0^{-2} \sum_{i,j} 
\boldsymbol{\nabla}_i \cdot \left(z+\tau_p^{-1}\right)^{-1}
\left(\left<\mathbf{f}_i \mathbf{f}_j\right>_{\text{lss}} 
- \left<\mathbf{f}_i\right>_{\text{lss}} \left<\mathbf{f}_j\right>_{\text{lss}}\right)
\cdot
\left[-\mathbf{F}_j^{\text{\text{eff}}}+\boldsymbol{\nabla}_j\right]
P_N(\mathbf{r}_1,...,\mathbf{r}_N;z).
\end{eqnarray}
\end{widetext}
where $\mathbf{F}_j^{\text{\text{eff}}}$ is the (normalized) effective force
acting on particle $j$ in the steady state,
\begin{equation}\label{Feff}
\mathbf{F}_j^{\text{\text{eff}}} = 
\boldsymbol{\nabla}_j \ln P_N^{\text{ss}}(\mathbf{r}_1, ..., \mathbf{r}_N).
\end{equation}
The right-hand-side of Eq. (\ref{proj2}) defines the effective 
evolution operator $\Omega^{\text{eff}}(z)$,
\begin{eqnarray}\label{Omegaeff}
\lefteqn{\Omega^{\text{eff}}(z) = \xi_0^{-2} \sum_{i,j} 
\boldsymbol{\nabla}_i \cdot \left(z+\tau_p^{-1}\right)^{-1} }
\nonumber \\ &&
\left(\left<\mathbf{f}_i \mathbf{f}_j\right>_{\text{lss}} 
- \left<\mathbf{f}_i\right>_{\text{lss}} \left<\mathbf{f}_j
\right>_{\text{lss}}\right)
\cdot
\left[-\mathbf{F}_j^{\text{\text{eff}}}+\boldsymbol{\nabla}_j\right].
\end{eqnarray} 
We recall that in approximation (\ref{appQOQ}) the influence of the correlations
 between self-propulsions and positions on the evolution of the self-propulsions
was neglected. In the effective evolution operator (\ref{Omegaeff}) the influence
of the same correlations on the (much slower) evolution of the particles'
positions is included through the steady-state correlations of the self-propulsions,
$\left<\mathbf{f}_i \mathbf{f}_j\right>_{\text{lss}} 
- \left<\mathbf{f}_i\right>_{\text{lss}} \left<\mathbf{f}_j
\right>_{\text{lss}}$.

Furthermore, 
we note that in the $z\to 0$ limit (which corresponds to coarse-graining in time) 
the evolution operator, Eq.~(\ref{Omegaeff}), becomes formally similar to the 
evolution operator for a system of Brownian particles with hydrodynamic interactions
\cite{Dhont}. This is the formal expression of the equivalence of a coarse-grained 
active system (\textit{i.e.} in the space of positions only and on long time 
scales) and a passive system.

Farage and Brader \cite{Farage} implicitly assumed the equivalence of a similarly
coarse-grained active system (\textit{i.e.} in the space of positions only and on long 
time scales) and a passive system \emph{without} hydrodynamic interactions. Specifically, 
in order to obtain the effective evolution operator used in Ref. \cite{Farage} 
from our operator (\ref{Omegaeff}), one needs to replace local steady-state correlations 
between the self-propulsions by the mean-square of the self-propulsion, 
$\left<\mathbf{f}_i \mathbf{f}_j\right>_{\text{lss}} 
- \left<\mathbf{f}_i\right>_{\text{lss}} \left<\mathbf{f}_j
\right>_{\text{lss}} \rightarrow \left<\mathbf{f}_i \mathbf{f}_i\right>_{\text{noise}}
\delta_{ij} \equiv D_f \tau_p \delta_{ij} \mathbf{I}$, where $\mathbf{I}$ is the unit 
tensor, and one needs to take the $z\to 0$ limit. 
Thus, our theory and the theory of Farage and Brader use different passive 
systems to approximate a coarse grained active system.

The most important assumption made in this section was approximating 
$\mathcal{Q}_{\text{lss}}\Omega\mathcal{Q}_{\text{lss}}$ by the free relaxation of
the self-propulsions. We shall see later that this assumption allows us to
recover the correct short time dependence of the intermediate scattering function. 

We recall once again that according to our model's evolution equations, Eqs. 
(\ref{eompos}-\ref{eomsp}), the self-propulsions evolve independently of the 
positions. However, in general, due to the coupled steady-state distribution of the 
self-propulsions and positions, the characteristic time for the transition from 
the short-time ballistic to diffusive motion (which is equal to $\tau_p$ for 
non-interacting particles, see Eq. (\ref{msdfree})) depends on the strength
of the interparticle interactions. We referred to this
change in Ref. \cite{activemct} as a renormalization of the relaxation
rate of the distribution of the self-propulsions by the interactions.
As we will see later, approximating 
$\mathcal{Q}_{\text{lss}}\Omega\mathcal{Q}_{\text{lss}}$ by the free relaxation of
the self-propulsions is equivalent to neglecting 
of any such renormalization. 

To improve upon this approximation one could try the alternative approach
presented in Appendix \ref{ap:alter}. This approach will likely lead to
a more accurate description of the short-time dynamics. However, this advantage
is out-weighted by its computational complexity. In the remainder of this paper
we will use the approximate effective evolution operator (\ref{Omegaeff}).

\section{Short-time dynamics: importance of 
correlations between self-propulsions and positions}\label{sec:corr}

Here we briefly discuss the short-time dynamics of the scattering functions.
First, we derive and discuss the exact expressions for the second time derivatives of
the scattering functions. We emphasize the importance of the correlations between
the self-propulsions acting on the particles and their positions for the correct
description of the short-time dynamics. Second, we point out that the effective 
evolution operator (\ref{Omegaeff}) leads to the correct
results for the second time derivatives of the scattering functions.

To evaluate the short-time behavior of the intermediate scattering
function we expand expression (\ref{Fqt}) in powers of $t$,
\begin{eqnarray}\label{Fqtst1}
F(q;t) &=&  \frac{1}{N}
\left<n(\mathbf{q}) n(-\mathbf{q})\right>
+ \frac{t}{N}
\left<n(\mathbf{q}) \Omega  n(-\mathbf{q})\right>
\nonumber \\ && + \frac{t^2}{2N}
\left<n(\mathbf{q}) \Omega^2 n(-\mathbf{q})\right> + ...
\end{eqnarray}
The first term at the right-hand-side of Eq. (\ref{Fqtst1}) is the steady
state static structure factor,
\begin{eqnarray}\label{Sq2}
S(q) = \frac{1}{N}
\left<n(\mathbf{q}) n(-\mathbf{q})\right>,
\end{eqnarray}
the second term vanishes due to symmetry, and the third term gives
\begin{eqnarray}\label{Fqtst2}
&& \frac{t^2}{2N}
\left<n(\mathbf{q}) \Omega^2  n(-\mathbf{q})\right> = 
\nonumber \\ \nonumber && 
- \frac{t^2 q^2}{2N\xi_0^2} \hat{\mathbf{q}}\cdot
\left<\sum_i \left(\mathbf{f}_i + \mathbf{F}_i \right)
e^{-i\mathbf{q}\cdot\mathbf{r}_i} 
\sum_j \left(\mathbf{f}_j + \mathbf{F}_j \right)
e^{i\mathbf{q}\cdot\mathbf{r}_j}\right>  \cdot \hat{\mathbf{q}}
\\ && = 
- \frac{t^2 q^2}{2} \omega_{\parallel}(q).
\end{eqnarray}

In Eq. (\ref{Fqtst2}), 
$\omega_{\parallel}(q)$ is a function that quantifies correlations of velocities
of individual particles,
\begin{eqnarray}\label{omegap}
&& \!\! \omega_{\parallel}(q) = \\ \nonumber &&
\frac{1}{N\xi_0^2}\hat{\mathbf{q}}\cdot
\left< \sum_{i,j}\left(\mathbf{f}_i + \mathbf{F}_i\right)
\left(\mathbf{f}_j + \mathbf{F}_j\right)
e^{-i\mathbf{q}\cdot\left(\mathbf{r}_i-\mathbf{r}_j\right)}\right>
\cdot\hat{\mathbf{q}}.
\end{eqnarray}
The interpretation of function $\omega_{\parallel}(q)$ comes from the fact that  
$\xi_0^{-1}\left(\mathbf{f}_i + \mathbf{F}_i\right)$
is the velocity of particle $i$ for our overdamped system, see Eq. (\ref{eompos}).
Computer simulation results \cite{activemct} show that,
at constant single particle effective temperature $T_{\text{eff}}$, 
Eq. (\ref{Teff}), with
increasing the persistence time (\textit{i.e.} with increasing departure from
equilibrium), $\omega_{\parallel}(q)$ gradually develops significant oscillations
while its overall magnitude decreases. 

To summarize, the short-time expansion reads,
\begin{eqnarray}\label{Fqtst3}
F(q;t) &=&  S(q)- \frac{q^2t^2}{2}\omega_{\parallel}(q)  + ...
\end{eqnarray}
This should be compared to the standard result for a colloidal suspension,
\begin{eqnarray}\label{FqtstBD}
F^{\text{BD}}(q;t) &=&  S(q)- q^2tD_0 H(q) + ...
\end{eqnarray}
where $D_0$ is the diffusion coefficient of a single isolated Brownian particle
and $H(q)$ is the hydrodynamic factor \cite{Dhont}
and the Newtonian dynamics result,
\begin{eqnarray}\label{FqtstND}
F^{\text{ND}}(q;t) &=&  S(q)- \frac{q^2t^2}{2} \frac{k_B T}{m}  + ...
\end{eqnarray}
where $k_B$ is the Boltzmann constant, $T$ is the temperature and 
$m$ is the particle mass.

The analogous calculation for the self-intermediate scattering function gives
\begin{eqnarray}\label{Fsqtst1}
F_s(q;t) &=& 1 - \frac{q^2t^2}{2}\omega_{\parallel}(\infty)  + ...
\end{eqnarray}
whereas for a colloidal system one gets
\begin{eqnarray}\label{FsqtstBD}
F_s^{\text{BD}}(q;t) &=& 1 - q^2tD_0 H(\infty)  + ...
\end{eqnarray}
where $D_0$ is the diffusion coefficient of a single isolated Brownian particle
and $H(\infty) = \lim_{q\to\infty} H(q)$, 
and the result for a Newtonian system reads  
\begin{eqnarray}\label{FsqtstND}
F_s^{\text{ND}}(q;t) &=&  1 - \frac{q^2t^2}{2} \frac{k_B T}{m}  + ...
\end{eqnarray}

The above results show that our active system has features of both colloidal and
Newtonian systems. Its short-time dynamics involves a non-trivial steady-state
correlation function $\omega_{\parallel}(q)$ quantifying spatial correlations
of particles' velocities which plays a role similar to that of the hydrodynamic
factor $H(q)$. On the other hand, the short-time dynamics of the active system 
is ballistic, like in a Newtonian system. However, the origin of this ballistic 
behavior is self-propulsion rather than inertia.

It can be showed that the same results for the short-time dynamics are obtained if
one starts from expressions for the scattering functions in terms of the effective
evolution operator (\ref{Omegaeff}). The difference between the exact expressions and 
those involving the effective evolution operator appears at $t^3$ order.

\section{Memory function representation}\label{sec:memf}

In this section we rewrite the formal expression (\ref{Fqt}) for the intermediate
scattering function in terms of the so-called frequency matrix
and irreducible memory matrix. The latter quantity contains all the (unknown)
non-trivial dynamic information about the system. The resulting expression for the
density correlation function in terms of the frequency matrix and the memory matrix 
is known as the memory function representation. 

We start by re-writing of the Laplace transform of the intermediate scattering function 
in terms of the approximate evolution operator (\ref{Omegaeff}),
\begin{eqnarray}\label{Fqz}
\lefteqn{\mathcal{LT}\left[F(q;t)\right](z) \equiv 
F(q;z)= } \nonumber \\ && 
N^{-1}\left<n(\mathbf{q})\left(z-\Omega\right)^{-1} n(-\mathbf{q})
\right> =  \nonumber \\ && N^{-1}\left<n(\mathbf{q}) 
\left(z-\Omega^{\text{eff}}(z)\right)^{-1} 
n(-\mathbf{q})\right>_\mathbf{r}.
\end{eqnarray}
Here $\left< ... \right>_\mathbf{r}$ denotes averaging over the 
steady-state distribution of particles' positions.

To derive the memory function representation of $F(q;z)$ we use the projection
operator approach \cite{Goetzebook,CHess,SL}. We define a projection operator 
on the microscopic density
\begin{equation}\label{Pn}
\mathcal{P}_n = ... \left. n(-\mathbf{q})\right>_\mathbf{r}
\left<n(\mathbf{q})n(-\mathbf{q})\right>_\mathbf{r}^{-1}
\left< n(\mathbf{q}) ... \right. .
\end{equation}
We emphasize that projection operator $\mathcal{P}_n$ is defined in terms
of the steady-state distribution, unlike in the approach of 
Farage and Brader \cite{Farage}. 
Next, we use the identity
\begin{eqnarray}\label{identity}
\frac{1}{z-\Omega^{\text{eff}}(z)} &=& \frac{1}{z - \Omega^{\text{eff}}(z)\mathcal{Q}_n} 
\\ \nonumber && + 
\frac{1}{z - \Omega^{\text{eff}}(z)\mathcal{Q}_n}\Omega^{\text{eff}}(z)\mathcal{P}_n 
\frac{1}{z-\Omega^{\text{eff}}(z)},
\end{eqnarray} 
where $\mathcal{Q}_n$ is the projection on the space orthogonal to that 
spanned by the microscopic density, 
\begin{equation}
\mathcal{Q}_n = \mathcal{I} - \mathcal{P}_n,
\end{equation}
to rewrite the Laplace transform of the time derivative of $N F(q;t)$ in the following 
way
\begin{widetext}
\begin{eqnarray}\label{timeder}
&& \mathcal{LT}[\partial_t N F(q;t)](z)= 
\left<n(\mathbf{q}) \Omega^{\text{eff}}(z) 
\frac{1}{z-\Omega^{\text{eff}}(z)} n(-\mathbf{q}) \right>_\mathbf{r} 
= \left<n(\mathbf{q}) \Omega^{\text{eff}}(z) \mathcal{P}_n 
\frac{1}{z-\Omega^{\text{eff}}(z)} n(-\mathbf{q}) \right>_\mathbf{r}
\nonumber \\ && + \left<n(\mathbf{q}) \Omega^{\text{eff}}(z) \mathcal{Q}_n 
\frac{1}{z-\Omega^{\text{eff}}(z)} n(-\mathbf{q}) \right>_\mathbf{r} 
=
\left<n(\mathbf{q}) \Omega^{\text{eff}}(z) n(-\mathbf{q})\right>_\mathbf{r}
\left<n(\mathbf{q})n(-\mathbf{q})\right>_\mathbf{r}^{-1}
\left< n(\mathbf{q}) \frac{1}{z-\Omega^{\text{eff}}(z)} n(-\mathbf{q}) 
\right>_\mathbf{r} 
\nonumber \\ && + 
\left<n(\mathbf{q}) \Omega^{\text{eff}}(z) \mathcal{Q}_n 
\frac{1}{z - \mathcal{Q}_n\Omega^{\text{eff}}(z)\mathcal{Q}_n} 
\mathcal{Q}_n \Omega^{\text{eff}}(z) n(-\mathbf{q})\right>_\mathbf{r}
\left<n(\mathbf{q})n(-\mathbf{q})\right>_\mathbf{r}^{-1}
\left< n(\mathbf{q})\frac{1}{z-\Omega^{\text{eff}}(z)}
n(-\mathbf{q}) \right>_\mathbf{r}.
\end{eqnarray}
\end{widetext}
The important part of the first term on the right-hand-side of the last equality sign in 
Eq. (\ref{timeder}) is the matrix element of the effective evolution operator, 
$\left<n(\mathbf{q}) \Omega^{\text{eff}}(z) n(-\mathbf{q})\right>_\mathbf{r}$, 
which can be expressed in terms of the frequency matrix $\mathcal{H}(q;z)$,
\begin{eqnarray}\label{freqmat1}
\left<n(\mathbf{q}) \Omega^{\text{eff}}(z) n(-\mathbf{q})\right>_\mathbf{r}
= -q^2 N \mathcal{H}(q;z).
\end{eqnarray}
The frequency matrix is given by the following expression
\begin{eqnarray}\label{freqmat2}
&& \mathcal{H}(q;z) = \\ \nonumber && 
\frac{ \hat{\mathbf{q}}\cdot
\left< \sum_{i,j}\left(\left<\mathbf{f}_i \mathbf{f}_j\right>_{\text{lss}} 
- \left<\mathbf{f}_i\right>_{\text{lss}} \left<\mathbf{f}_j\right>_{\text{lss}}\right) 
e^{-i\mathbf{q}\cdot\left(\mathbf{r}_i-\mathbf{r}_j\right)}\right>_\mathbf{r} 
\cdot\hat{\mathbf{q}}}
{N\xi_0^2\left(z+\tau_p^{-1}\right)}.
\end{eqnarray}
In Eq. (\ref{freqmat2}) $\hat{\mathbf{q}}$ is a unit vector, 
$\hat{\mathbf{q}}=\mathbf{q}/q$.
In turn, the non-trivial part of the frequency matrix is represented by the function 
$\omega_{\parallel}(q)$ introduced in Sec. \ref{sec:corr}, which 
quantifies correlations of velocities of individual particles,
\begin{eqnarray}\label{omegap2}
&& \!\!\!\!    \frac{1}{N\xi_0^2}
\hat{\mathbf{q}}\!\cdot\!
\left< \sum_{i,j}\left(\left<\mathbf{f}_i \mathbf{f}_j\right>_{\text{lss}} 
- \left<\mathbf{f}_i\right>_{\text{lss}} \left<\mathbf{f}_j\right>_{\text{lss}}\right) 
e^{-i\mathbf{q}\cdot\left(\mathbf{r}_i-\mathbf{r}_j\right)}\right>_\mathbf{r}
\!\!\!\cdot\!\hat{\mathbf{q}} = 
\nonumber \\ && \!\!\!\! 
\frac{1}{N\xi_0^2}\hat{\mathbf{q}}\!\cdot\!
\left< \sum_{i,j}\left(\left< \mathbf{f}_i\right>_{\text{lss}} + \mathbf{F}_i\right)
\left(\left< \mathbf{f}_j\right>_{\text{lss}} + \mathbf{F}_j\right)
e^{-i\mathbf{q}\cdot\left(\mathbf{r}_i-\mathbf{r}_j\right)}\right>_{\mathbf{r}}
\!\!\!\cdot\!\hat{\mathbf{q}} =
\nonumber \\ && \!\!\!\!
\frac{1}{N\xi_0^2}\hat{\mathbf{q}}\!\cdot\!
\left< \sum_{i,j}\left(\mathbf{f}_i + \mathbf{F}_i\right)
\left(\mathbf{f}_j + \mathbf{F}_j\right)
e^{-i\mathbf{q}\cdot\left(\mathbf{r}_i-\mathbf{r}_j\right)}\right>
\!\!\!\cdot\!\hat{\mathbf{q}} \equiv \omega_{\parallel}(q)
\nonumber \\ 
\end{eqnarray}
To arrive at the penultimate line of Eq. (\ref{omegap}) we utilized Eq. (\ref{avsp}) 
and to arrive at the last line we used  
the fact that the local steady-state average followed by averaging over
particles' positions is equivalent to the full steady-state average,
\begin{equation}
\left< \left< ... \right>_{\text{lss}}\right>_{\mathbf{r}}=\left< ... \right>.
\end{equation}
We note that the penultimate line of Eq. (\ref{omegap}) shows that 
$\omega_{\parallel}(q)$ quantifies fluctuations around the force balance 
condition (\ref{avsp}). 

The important part of the second term at the right-hand-side of the last equality 
sign in Eq. (\ref{timeder}) can be expressed in terms of reducible \cite{CHess,Kawasaki} 
memory matrix $\mathcal{M}(q;z)$,
\begin{eqnarray}\label{memfunred1}
&& \left<n(\mathbf{q}) \Omega^{\text{eff}}(z) \mathcal{Q}_n 
\frac{1}{z - \mathcal{Q}_n\Omega^{\text{eff}}(z)\mathcal{Q}_n} 
\mathcal{Q}_n \Omega^{\text{eff}}(z) n(-\mathbf{q})\right>_{\mathbf{r}} =
\nonumber \\ && q^2 N \mathcal{M}(q;z).
\end{eqnarray}
The memory matrix is given by the following expression
\begin{widetext}
\begin{eqnarray}\label{memfunred2}
&&  \mathcal{M}(q;z) = 
\left(N \xi_0^4\left(z+\tau_p^{-1}\right)^2\right)^{-1}
\hat{\mathbf{q}}\cdot\left< \sum_{i,j} e^{-i\mathbf{q}\cdot\mathbf{r}_i}
\left(\left<\mathbf{f}_i \mathbf{f}_j\right>_{\text{lss}} 
- \left<\mathbf{f}_i\right>_{\text{lss}} 
\left<\mathbf{f}_j\right>_{\text{lss}}\right)\cdot
\right. \nonumber \\ && \times \left.
\left[-\boldsymbol{\nabla}_j + \mathbf{F}^{\text{eff}}_j\right]
\mathcal{Q}_n
\frac{1}{z - \mathcal{Q}_n\Omega^{\text{eff}}(z)\mathcal{Q}_n}
\mathcal{Q}_n \sum_{l,m} \boldsymbol{\nabla}_l \cdot
\left(\left<\mathbf{f}_l \mathbf{f}_m\right>_{\text{lss}} 
- \left<\mathbf{f}_l\right>_{\text{lss}} \left<\mathbf{f}_m\right>_{\text{lss}}\right)
e^{i\mathbf{q}\cdot\mathbf{r}_m} \right>_{\mathbf{r}}\cdot\hat{\mathbf{q}}
\end{eqnarray}
\end{widetext}

We can now rewrite the Laplace transform of the intermediate
scattering function in terms of the frequency and memory matrix,
\begin{eqnarray}\label{fqzmfred}
F(q;z) = \frac{S(q)}{z+q^2\left(\mathcal{H}(q;z) - \mathcal{M}(q;z)\right)/S(q)} 
\end{eqnarray}
where $S(q)$ is the steady-state structure factor,
\begin{eqnarray}\label{Sq}
S(q)= \left<n(\mathbf{q})n(-\mathbf{q})\right>_\mathbf{r} 
\equiv \left<n(\mathbf{q})n(-\mathbf{q})\right>.
\end{eqnarray}
The second equality sign in Eq.~(\ref{Sq}) follows from the fact that for 
self-propulsion-independent quantities averaging over particles' positions 
is equivalent to averaging over the full steady-state distribution of
positions and self-propulsions. 

Eq. (\ref{fqzmfred}) is the active matter equivalent of the memory
function representation of the intermediate scattering function of a Brownian
colloidal suspension derived by Ackerson \cite{Ackerson}. As pointed out
by Cichocki and Hess \cite{CHess} and later elaborated by Kawasaki \cite{Kawasaki},
the memory matrix that enters into Ackerson's formula is not fully irreducible. 
Both Cichocki and Hess, and Kawasaki argued that approximations should be
applied to the fully irreducible memory matrix. Therefore, we 
will follow the analysis presented in Refs. \cite{CHess,Kawasaki} and 
derive an irreducible memory matrix. 

We define the irreducible evolution operator $\Omega^{\mathrm{irr}}(z)$,
\begin{equation}\label{Omegairr1}
\Omega^{\text{irr}}(z) = \mathcal{Q}_n\Omega^{\text{eff}}(z)\mathcal{Q}_n -
\delta \Omega^{\text{irr}}(z)
\end{equation}
where the subtraction term  $\delta \Omega^{\text{irr}}(z)$ \cite{commentlong} reads 
\begin{widetext}
\begin{eqnarray}\label{Omegairr2}
\delta \Omega^{\text{irr}}(z) &=& \left. \mathcal{Q}_n
\left(N \xi_0^4\left(z+\tau_p^{-1}\right)^2\right)^{-1}\sum_{i,j}
\boldsymbol{\nabla}_i \cdot \left(\left<\mathbf{f}_i \mathbf{f}_j\right>_{\text{lss}} 
- \left<\mathbf{f}_i\right>_{\text{lss}} \left<\mathbf{f}_j\right>_{\text{lss}}\right)
e^{i\mathbf{q}\cdot \mathbf{r}_j}\right>_{\mathbf{r}}\cdot\hat{\mathbf{q}}
\left(\mathcal{H}(q;z)\right)^{-1}
\nonumber \\ && \hat{\mathbf{q}}\cdot
\left< \sum_{k,l} e^{-i\mathbf{q}\cdot \mathbf{r}_k}
\left(\left<\mathbf{f}_k \mathbf{f}_l\right>_{\text{lss}} 
- \left<\mathbf{f}_k\right>_{\text{lss}} \left<\mathbf{f}_l\right>_{\text{lss}}\right)
\cdot \left(\boldsymbol{\nabla}_l-\beta\mathbf{F}^{\text{eff}}_l\right)\mathcal{Q}_n 
\right.
\end{eqnarray}
Next, we define the irreducible memory matrix 
$\mathcal{M}^{\text{irr}}(q;z)$, which is given by the expression analogous to 
Eq. (\ref{memfunred2}) but with the projected evolution operator
$\mathcal{Q}_n\Omega^{\text{eff}}(z)\mathcal{Q}_n$ replaced by 
irreducible evolution  operator $\Omega^{\text{irr}}(z)$,
\begin{eqnarray}\label{memfunirr1}
&&  \mathcal{M}^{\text{irr}}(q;z) = 
\left(N \xi_0^4\left(z+\tau_p^{-1}\right)^2\right)^{-1}
\hat{\mathbf{q}}\cdot\left< \sum_{i,j} e^{-i\mathbf{q}\cdot\mathbf{r}_i}
\left(\left<\mathbf{f}_i \mathbf{f}_j\right>_{\text{lss}} 
- \left<\mathbf{f}_i\right>_{\text{lss}} 
\left<\mathbf{f}_j\right>_{\text{lss}}\right)\cdot
\right. \nonumber \\ && \times \left.
\left[-\boldsymbol{\nabla}_j + \mathbf{F}^{\text{eff}}_j\right]
\mathcal{Q}_n
\frac{1}{z - \Omega^{\text{irr}}(z)}
\mathcal{Q}_n \sum_{l,m} \boldsymbol{\nabla}_l \cdot
\left(\left<\mathbf{f}_l \mathbf{f}_m\right>_{\text{lss}} 
- \left<\mathbf{f}_l\right>_{\text{lss}} \left<\mathbf{f}_m\right>_{\text{lss}}\right)
e^{i\mathbf{q}\cdot\mathbf{r}_m} \right>_{\mathbf{r}}\cdot\hat{\mathbf{q}}.
\end{eqnarray}
\end{widetext}
Finally, we use an identity similar to Eq. (\ref{identity}),
\begin{eqnarray}\label{identity2}
\lefteqn{\frac{1}{z-\mathcal{Q}_n\Omega^{\text{eff}}(z)\mathcal{Q}_n} 
= \frac{1}{z - \Omega^{\text{irr}}(z)} }
\\ \nonumber && + 
\frac{1}{z - \Omega^{\text{irr}}(z)}\delta\Omega^{\text{eff}}(z) 
\frac{1}{z-\mathcal{Q}_n\Omega^{\text{eff}}(z)\mathcal{Q}_n},
\end{eqnarray} 
and we derive the following relation between $\mathcal{M}(q;z)$ and 
$\mathcal{M}^{\text{irr}}(q;z)$,
\begin{equation}\label{redirr}
\mathcal{M}(q;z)=\mathcal{M}^{\text{irr}}(q;z)-\mathcal{M}^{\text{irr}}(q;z)
\mathcal{H}^{-1}(q;z)\mathcal{M}(q;z).
\end{equation}

Combining Eqs. (\ref{fqzmfred}) and (\ref{redirr}) we arrive at the following
representation of the intermediate scattering function in terms of the
irreducible memory matrix,
\begin{eqnarray}\label{fqzmfirr}
F(q;z) = \frac{S(q)}{z+\frac{q^2\mathcal{H}(q;z)/S(q)}
{1+\mathcal{M}^{\text{irr}}(q;z)/\mathcal{H}(q;z)}}.
\end{eqnarray}

Eq. (\ref{fqzmfirr}) is the active matter equivalent of the memory
function representation of the intermediate scattering function of a Brownian
colloidal suspension derived by Cichocki and Hess \cite{CHess}. The latter
equation was the starting point of the derivation of the mode-coupling
theory for the glassy dynamics of colloidal systems \cite{SL}. 

In spite of the formal similarity between Eq. (\ref{fqzmfirr}) and the irreducible memory
function representation of the intermediate scattering function of a Brownian system, 
these equations are only equivalent if we take the limit $z\to 0$ in the frequency 
matrix term. In fact, as we have already stated a couple of times, active
system performs interacting \emph{persistent} Brownian motion and, therefore, its short
time dynamics is ballistic rather than diffusive  (we emphasize that, in this context, 
the term ballistic does not imply inertia, which is irrelevant for our overdamped 
system). The ballistic character of the short-time 
dynamics can be clearly seen if Eq. (\ref{fqzmfirr}) is 
re-written in the following way. First, we rewrite Eq. (\ref{fqzmfirr}) by introducing 
the irreducible \emph{memory function}  $M^{\text{irr}}(q;z)$,
\begin{eqnarray}\label{fqzmfirr2}
&& \left(z+\tau_p^{-1}
+ M^{\text{irr}}(q;z)\right)
\left(z F(q;z) - F(q;t=0)\right) 
\nonumber \\ && = -\left(\omega_{\parallel}(q) q^2/S(q)\right)F(q;z).
\end{eqnarray}
where the memory function reads
\begin{eqnarray}\label{memfunirr2}
M^{\text{irr}}(q;z) = 
\left(z+\tau_p^{-1}\right)^2\mathcal{M}^{\text{irr}}(q;z)/\omega_{\parallel}(q).
\end{eqnarray}
Next, we rewrite Eq. (\ref{fqzmfirr2}) in the time domain,
\begin{eqnarray}\label{memfunirr3}
&& \partial^2_t F(q;t) +\tau_p^{-1} \partial_t  F(q;t) 
+ \frac{\omega_{\parallel}(q) q^2}{S(q)} F(q;t) =  
\nonumber \\  &&  
- \int_0^t dt' M^{\text{irr}}(q;t-t') \partial_{t'} F(q;t'),
\end{eqnarray}
where $M^{\text{irr}}(q;t)$ is the inverse Laplace transform of
$M^{\text{irr}}(q;z)$. We note that the resulting equation, Eq. (\ref{memfunirr3}), 
has the same form as the memory function
equation for the intermediate scattering function of a Newtonian system. 

Let us discuss the meaning of the different terms in Eq. (\ref{memfunirr3}), 
starting with the three terms at its left-hand-side. 
The presence of the second time derivative implies 
ballistic dynamics at short times. For our system of self-propelled particles 
the origin of the ballistic behavior is the persistence of the microscopic motion. 
The first time derivative term describes the relaxation
of the ballistic motion due to the evolution of the self-propulsion. Since
we neglected the influence of the interactions on the evolution of the 
self-propulsion (recall Eq. (\ref{appQOQ})), the relaxation time of the self-propulsion
is unchanged and equal to $\tau_p$. The third term at the left-hand-side of 
Eq. (\ref{memfunirr3}) describes the collective random motion of the 
particles on time scale longer than the persistence time of the self-propulsion. 
On such a time scale, the second derivative term in Eq. (\ref{memfunirr3})
can be neglected. Then (recall that we are neglecting the right-hand side term
for a moment), the intermediate scattering function
$F(q;t)$ evolves diffusively, 
\begin{equation}\label{FqtnoM}
F(q;t) \propto \exp\left(-\frac{\omega_{\parallel}(q) \tau_p q^2}{S(q)}t\right). 
\end{equation}
and $\omega_{\parallel}(q) \tau_p/S(q)$ plays the role of the 
short-time collective diffusion coefficient. 
This simple exponential time dependence of density correlations is modified 
by the presence of the memory function term at the right-hand-side
of Eq. (\ref{memfunirr3}). The memory function term describes an internal time-delayed 
friction generated by the interparticle interactions. This 
term leads to slow and glassy dynamics through a feedback mechanism implicit
in the mode-coupling approximation (see discussion after Eq. (\ref{memfction})).

The non-equilibrium nature of the active system manifests itself in  
Eq. (\ref{memfunirr3}) through the presence of 
$\omega_\parallel(q)$, which quantifies spatial correlations between velocities
of different particles. In particular, the second time derivative of the intermediate 
scattering function at $t=0$ is expressed in
terms of $\omega_\parallel(q)$,
\begin{equation}\label{tderFqt}
\left. \partial^2_t F(q;t)\right|_{t=0} 
= -\frac{\omega_{\parallel}(q) q^2}{S(q)} F(q;t=0) 
\equiv -\omega_{\parallel}(q) q^2,
\end{equation}
Eq. (\ref{tderFqt}) agrees with the short-time expansion discussed in Sec. 
\ref{sec:corr}. In the next section we will show that
$\omega_\parallel(q)$ also enters an approximate expression for the 
so-called vertex function. 

\section{Mode-coupling-like approximation}\label{sec:mct}

In order to make Eq. (\ref{memfunirr3}) useful we need an explicit expression
for the memory function. Since the memory function contains all the non-trivial dynamic 
information about the system, there is little hope to derive an exact 
expression for it. Here we use the factorization approximation, which is also
at the heart of the mode-coupling theory of glassy dynamics, \cite{Goetzebook} and thus
we call our approach a mode-coupling-like approximation. 

Specifically, to derive an approximate expression for the memory function 
we follow the steps of the derivation of the mode-coupling theory for systems
evolving with Brownian dynamics \cite{SL}. The derivation consists of three steps. 

First, we project onto the subspace of density pairs,
\begin{widetext}
\begin{eqnarray}\label{projdenpairs}
\lefteqn{\hat{\mathbf{q}}\cdot\left< \sum_{i,j} e^{-i\mathbf{q}\cdot\mathbf{r}_i}
\left(\left<\mathbf{f}_i \mathbf{f}_j\right>_{\text{lss}} 
- \left<\mathbf{f}_i\right>_{\text{lss}} 
\left<\mathbf{f}_j\right>_{\text{lss}}\right)\cdot
\left[-\boldsymbol{\nabla}_j + \mathbf{F}^{\text{eff}}_j\right]
\mathcal{Q}_n \right. }
\nonumber \\ \nonumber && \approx \sum_{\mathbf{q}_1,...,\mathbf{q}_4}
\hat{\mathbf{q}}\cdot\left< \sum_{i,j} e^{-i\mathbf{q}\cdot\mathbf{r}_i}
\left(\left<\mathbf{f}_i \mathbf{f}_j\right>_{\text{lss}} 
- \left<\mathbf{f}_i\right>_{\text{lss}} 
\left<\mathbf{f}_j\right>_{\text{lss}}\right)\cdot
\left[-\boldsymbol{\nabla}_j + \mathbf{F}^{\text{eff}}_j\right]
\mathcal{Q}_n n_2(-\mathbf{q}_1,-\mathbf{q}_2)\right>_{\mathbf{r}} 
\\ && \times 
\left[\left<\mathcal{Q}_n n_2(\mathbf{q}_1,\mathbf{q}_2) \mathcal{Q}_n 
n_2(-\mathbf{q}_3,-\mathbf{q}_4)
\right>_{\mathbf{r}}\right]^{-1}\left< \mathcal{Q}_n n_2(\mathbf{q}_3,\mathbf{q}_4)
\right. .
\end{eqnarray}
\end{widetext}
Here $n_2(\mathbf{q}_1,\mathbf{q}_2)$ is the Fourier transform of the microscopic 
two-particle density,
\begin{equation}\label{n2def}
n_2(\mathbf{q}_1,\mathbf{q}_2) = \sum_{l,m} 
e^{-i\mathbf{q}_1\cdot\mathbf{r}_l-i\mathbf{q}_2\cdot\mathbf{r}_m},
\end{equation}
and $\left[\left<\mathcal{Q}_n n_2(\mathbf{q}_1,\mathbf{q}_2) \mathcal{Q}_n 
n_2(-\mathbf{q}_3,-\mathbf{q}_4)
\right>_{\mathbf{r}}\right]^{-1}$ is the inverse of the correlation matrix of 
microscopic pair densities. 
We should emphasize the importance of using in Eq. (\ref{projdenpairs}) 
only the parts of the microscopic pair density that are orthogonal to the 
microscopic density, \textit{i.e.} using $\mathcal{Q}_n n_2(\mathbf{q}_1,\mathbf{q}_2)$
rather than $n_2(\mathbf{q}_1,\mathbf{q}_2)$. The presence
of the operator $\mathcal{Q}_n$ is necessary for the existence of the inverse
of the correlation matrix 
$\left<\mathcal{Q}_n n_2(\mathbf{q}_1,\mathbf{q}_2) \mathcal{Q}_n 
n_2(-\mathbf{q}_3,-\mathbf{q}_4)\right>_{\mathbf{r}}$ \cite{Anderson}. 

Second, we factorize averages resulting from substituting projection
(\ref{projdenpairs}) into the expression for the memory function and
\textit{at the same time} replace the irreducible operator $\Omega^{\text{irr}}(z)$
by effective evolution operator $\Omega^{\text{eff}}(z)$. We should emphasize that 
this factorization has to be done in the time domain,
\begin{widetext}
\begin{eqnarray}\label{fac4to2}
\lefteqn{\mathcal{LT}^{-1}\left[\left< \mathcal{Q}_n n_2(\mathbf{q}_1,\mathbf{q}_2) 
\left(z-\Omega^{\text{irr}}(z)\right)^{-1} \mathcal{Q}_n 
n_2(-\mathbf{q}_3,-\mathbf{q}_4)\right>_{\mathbf{r}}\right] \approx}
\\ \nonumber && 
\mathcal{LT}^{-1}\left[\left< n(\mathbf{q}_1) 
\left(z-\Omega^{\text{eff}}(z)\right)^{-1} n(-\mathbf{q}_3)\right>_{\mathbf{r}}\right]
\mathcal{LT}^{-1}\left[\left< n(\mathbf{q}_2) 
\left(z-\Omega^{\text{eff}}(z)\right)^{-1} n(-\mathbf{q}_4)\right>_{\mathbf{r}}\right]
+ \left\{ 3 \leftrightarrow 4 \right\}.
\end{eqnarray}
\end{widetext}
Here $\mathcal{LT}^{-1}$ denotes the inverse Laplace transform and
$\left\{ 3 \leftrightarrow 4 \right\}$ means the preceding expression with
labels $3$ and $4$ interchanged. Consistently with
Eq. (\ref{fac4to2}) we also factorize the steady-state correlation matrix
of microscopic pair densities and for its inverse we get
\begin{eqnarray}\label{fac4to2ss}
&& \left[\left<\mathcal{Q}_n n_2(\mathbf{q}_1,\mathbf{q}_2) \mathcal{Q}_n 
n_2(-\mathbf{q}_3,-\mathbf{q}_4)
\right>_{\mathbf{r}}\right]^{-1} \approx
\\ \nonumber && 
\left<n(\mathbf{q}_1) n(-\mathbf{q}_3)\right>_{\mathbf{r}}^{-1}
\left<n(\mathbf{q}_2) n(-\mathbf{q}_4)\right>_{\mathbf{r}}^{-1}
+ \left\{ 3 \leftrightarrow 4 \right\}.
\end{eqnarray}

Third, we approximate the vertex functions. Due to the presence of the velocity 
correlations, this last step is somewhat more complex 
than the approximation used in the derivation of the standard mode-coupling theory 
\cite{SL}. We will explain it on the 
example of the left vertex, $\mathcal{V}_l$, which is given by the following formula
\begin{eqnarray}\label{leftvertexdef}
\lefteqn{ \mathcal{V}_l(\mathbf{q};\mathbf{q}_1,\mathbf{q}_2) = }
\nonumber \\ && 
\xi_0^{-2} \hat{\mathbf{q}}\cdot\left< \sum_{i,j} e^{-i\mathbf{q}\cdot\mathbf{r}_i}
\left(\left<\mathbf{f}_i \mathbf{f}_j\right>_{\text{lss}} 
- \left<\mathbf{f}_i\right>_{\text{lss}} \left<\mathbf{f}_j\right>_{\text{lss}}\right)
\cdot \right.
\nonumber \\ && \times \left.
\left[-\nabla_j + \mathbf{F}^{\text{eff}}_j\right]
\mathcal{Q}_n n_2(-\mathbf{q}_1,-\mathbf{q}_2)\right>_{\mathbf{r}}
\nonumber \\ &=&
\xi_0^{-2} \hat{\mathbf{q}}\cdot\left< \sum_{i,j} e^{-i\mathbf{q}\cdot\mathbf{r}_i}
\left(\mathbf{f}_i + \mathbf{F}_i\right)
\left(\mathbf{f}_j + \mathbf{F}_j\right)\cdot \right.
\nonumber \\ && \times \left.
\left[-\nabla_j + \mathbf{F}^{\text{eff}}_j\right]
\mathcal{Q}_n n_2(-\mathbf{q}_1,-\mathbf{q}_2)\right>
\end{eqnarray}
Due to the presence of the projection operator $\mathcal{Q}_n$, 
the left vertex consists of two terms,
\begin{eqnarray}\label{leftvertex1}
\lefteqn{ \mathcal{V}_l(\mathbf{q};\mathbf{q}_1,\mathbf{q}_2) = }
\\ \nonumber 
&-& \xi_0^{-2} \hat{\mathbf{q}}\cdot\left< \sum_{i,j} e^{-i\mathbf{q}\cdot\mathbf{r}_i}
\left(\mathbf{f}_i + \mathbf{F}_i\right)
\left(\mathbf{f}_j + \mathbf{F}_j\right)\cdot \right.
\\ \nonumber && \times \left.
\left[\nabla_j n_2(-\mathbf{q}_1,-\mathbf{q}_2)\right]\right>
\\ \nonumber 
&+& \xi_0^{-2} \hat{\mathbf{q}}\cdot\left< \sum_{i,j} e^{-i\mathbf{q}\cdot\mathbf{r}_i}
\left(\mathbf{f}_i + \mathbf{F}_i\right)
\left(\mathbf{f}_j + \mathbf{F}_j\right)\cdot \right.
\\ \nonumber && \times \left.
\left[\nabla_j n(-\mathbf{q})\right]\right>
\left<n(\mathbf{q})n(-\mathbf{q})\right>^{-1}
\left<n(\mathbf{q})n_2(-\mathbf{q}_1,-\mathbf{q}_2)\right>.
\end{eqnarray}

We will start with the second term, which is a bit easier to analyze. With the help of
convolution approximation \cite{HansenMcDonald} generalized to the stationary 
state of our active system,
\begin{eqnarray}\label{convolution}
&& \left< n(\mathbf{q}) n_2(-\mathbf{q}_1,-\mathbf{q}_2)\right> = 
\left< \sum_i \sum_{j,k} 
e^{-i\mathbf{q}\cdot\mathbf{r}_i}e^{i\mathbf{q}_1\cdot\mathbf{r}_j}  
e^{i\mathbf{q}_2\cdot\mathbf{r}_k} \right>
\nonumber \\ && \approx 
N S(q)S(q_1)S(q_2)\delta_{\mathbf{q},\mathbf{q}_1+\mathbf{q}_2},
\end{eqnarray}
we can rewrite the second term in the following form,
\begin{eqnarray}\label{leftvertex2}
&&\xi_0^{-2} \hat{\mathbf{q}}\cdot\left< \sum_{i,j} e^{-i\mathbf{q}\cdot\mathbf{r}_i}
\left(\mathbf{f}_i + \mathbf{F}_i\right)
\left(\mathbf{f}_j + \mathbf{F}_j\right)\cdot \right.
\nonumber \\ \nonumber && \times \left.
\left[\nabla_j n(-\mathbf{q})\right]\right>
\left<n(\mathbf{q})n(-\mathbf{q})\right>^{-1}
\left<n(\mathbf{q})n_2(-\mathbf{q}_1,-\mathbf{q}_2)\right>
\\ \nonumber & \approx &
i \xi_0^{-2} \hat{\mathbf{q}}\cdot 
\left< \sum_{i,j}\left(\mathbf{f}_i + \mathbf{F}_i\right)
\left(\mathbf{f}_j + \mathbf{F}_j\right)
e^{-i\mathbf{q}\cdot\left(\mathbf{r}_i-\mathbf{r}_j\right)}\right>
\cdot\mathbf{q}
\\ \nonumber && \times 
S(q_1)S(q_2)\delta_{\mathbf{q},\mathbf{q}_1+\mathbf{q}_2}
\\ & = &
i N \omega_{\parallel}(q) \hat{\mathbf{q}}\cdot \left(\mathbf{q}_1+\mathbf{q}_2\right)
S(q_1)S(q_2)\delta_{\mathbf{q},\mathbf{q}_1+\mathbf{q}_2},
\end{eqnarray}
where in the last step we used the fact that for an isotropic system the tensor
$\left< \sum_{i,j}\left(\mathbf{f}_i + \mathbf{F}_i\right)
\left(\mathbf{f}_j + \mathbf{F}_j\right)
e^{-i\mathbf{q}\cdot\left(\mathbf{r}_i-\mathbf{r}_j\right)}\right>$
has only components parallel and perpendicular to $\hat{\mathbf{q}}$, and that 
its parallel component is proportional to $\omega_{\parallel}(q)$, see 
Eq. (\ref{omegap}).

The first term at the right-hand-side of Eq. (\ref{leftvertex1}) can be
rewritten as follows,
\begin{eqnarray}\label{leftvertex3}
&&
-\xi_0^{-2} \hat{\mathbf{q}}\cdot\left< \sum_{i,j} e^{-i\mathbf{q}\cdot\mathbf{r}_i}
\left(\mathbf{f}_i + \mathbf{F}_i\right)
\left(\mathbf{f}_j + \mathbf{F}_j\right)\cdot \right.
\\ \nonumber && \times \left.
\left[\nabla_j n_2(-\mathbf{q}_1,-\mathbf{q}_2)\right]\right>
\\ \nonumber & = & 
-\xi_0^{-2} \hat{\mathbf{q}}\cdot\left< \sum_{i,j} e^{-i\mathbf{q}\cdot\mathbf{r}_i}
\left(\mathbf{f}_i + \mathbf{F}_i\right)
\left(\mathbf{f}_j + \mathbf{F}_j\right)\cdot \right.
\\ \nonumber && \times \left. 
\left[i\mathbf{q}_1 e^{i\mathbf{q}_1\cdot\mathbf{r}_j}\sum_l  
e^{i\mathbf{q}_2\cdot\mathbf{r}_l} + \left\{1\leftrightarrow 2\right\} \right]\right>.
\end{eqnarray}
Now, we use an approximation which is a generalization of the convolution
approximation to correlation functions involving active particles' velocities,
\begin{eqnarray}\label{convolutiona}
&& \xi_0^{-2}\left< \sum_{i,j} e^{-i\mathbf{q}\cdot\mathbf{r}_i}
\left(\mathbf{f}_i + \mathbf{F}_i\right)
\left(\mathbf{f}_j + \mathbf{F}_j\right)e^{i\mathbf{q}_1\cdot\mathbf{r}_j}
\sum_l e^{i\mathbf{q}_2\cdot\mathbf{r}_l}\right> \approx
\nonumber \\ && 
N \xi_0^2 \boldsymbol{\omega}(\mathbf{q})
\cdot \left(\left<\delta\mathbf{f} \delta\mathbf{f} \right>\right)^{-1} \cdot
\boldsymbol{\omega}(\mathbf{q}_1)S(q_2)
\delta_{\mathbf{q},\mathbf{q}_1+\mathbf{q}_2}.
\end{eqnarray}
Here $\boldsymbol{\omega}(\mathbf{q})$ is the tensorial version of 
$\omega_{\parallel}(q)$,
\begin{eqnarray}\label{omegaten}
\boldsymbol{\omega}(\mathbf{q})= \frac{1}{N\xi_0^2}
\left< \sum_{i,j}\left(\mathbf{f}_i + \mathbf{F}_i\right)
\left(\mathbf{f}_j + \mathbf{F}_j\right)
e^{-i\mathbf{q}\cdot\left(\mathbf{r}_i-\mathbf{r}_j\right)}\right>,
\end{eqnarray}
and $\left(\left<\delta\mathbf{f} \delta\mathbf{f} \right>\right)^{-1}$ is
the inverse of the correlation matrix of single-particle velocities, 
\begin{eqnarray}\label{ff}
\left<\delta\mathbf{f} \delta\mathbf{f} \right> = 
\frac{1}{N}\left< \sum_i \left(\mathbf{f}_i + \mathbf{F}_i\right)
\left(\mathbf{f}_i + \mathbf{F}_i\right)\right>.
\end{eqnarray}
Note that $\lim_{q\to\infty} \boldsymbol{\omega}(\mathbf{q}) = 
\left<\delta\mathbf{f} \delta\mathbf{f} \right>/\xi_0^2$.
Using convolution approximation (\ref{convolutiona}) and symmetry properties
of tensors $\boldsymbol{\omega}(\mathbf{q})$ and 
$\left<\delta\mathbf{f} \delta\mathbf{f} \right>$
we can finally write the first term (\ref{leftvertex3}) in the following form
\begin{eqnarray}\label{leftvertex4}
&& - iN\hat{\mathbf{q}}\cdot\left[\mathbf{q}_1 \omega_{\parallel}(q)
\omega_{\parallel}(q_1)S(k_2)/\omega_{\parallel}(\infty)\right.
\nonumber \\ && 
\left. + \mathbf{q}_2 \omega_{\parallel}(q)
\omega_{\parallel}(q_2)S(q_1)/\omega_{\parallel}(\infty)\right]
\delta_{\mathbf{q},\mathbf{q}_1+\mathbf{q}_2},
\end{eqnarray}
where $\omega_{\parallel}(\infty)=\lim_{q\to\infty}\omega_{\parallel}(q)
\equiv \left(3N\xi_0^2\right)^{-1} 
\left<\sum_i \left(\mathbf{f}_i+\mathbf{F}_i\right)^2 \right>$.

Combining Eqs. (\ref{leftvertex2}) and (\ref{leftvertex4}) we get the 
following approximate expression for the left vertex,
\begin{eqnarray}\label{leftvertex5}
\lefteqn{ \mathcal{V}_l(\mathbf{q};\mathbf{q}_1,\mathbf{q}_2) 
\approx -i N S(q_1)S(q_2) \omega_{\parallel}(q)}
\nonumber \\ & \times & 
\left[ \hat{\mathbf{q}}\cdot\mathbf{q}_1 \left(
\frac{\omega_{\parallel}(q_1)}{\omega_{\parallel}(\infty)S(q_1)} - 1\right)
+ \left\{ 1\leftrightarrow 2 \right\} \right]
\nonumber \\ & \equiv & 
i N \rho S(q_1)S(q_2) \omega_{\parallel}(q)
\left[ \hat{\mathbf{q}}\cdot\mathbf{q}_1 \mathcal{C}(q_1) +
\hat{\mathbf{q}}\cdot\mathbf{q}_2 \mathcal{C}(q_2)\right].
\end{eqnarray}
where a new function $\mathcal{C}(q)$ reads
\begin{eqnarray}\label{newc}
\rho \mathcal{C}(q) = 1-\frac{\omega_{\parallel}(q)}
{\omega_{\parallel}(\infty )S(q)}.
\end{eqnarray}
We note that if correlations between particles' velocities are neglected 
(\textit{i.e.} if $\omega_{\parallel}(q)$ is set equal to 1), $\mathcal{C}(q)$
becomes an analogue of the direct correlation function for our steady-state 
active system. 

The right vertex can be analyzed in the same way. We note that in the standard
mode-coupling theory there is a close correspondence between the vertices that
enter the expression for the memory function for the collective intermediate 
scattering function and those that enter the expression for the memory function 
for the self-intermediate scattering function. We should emphasize that this 
correspondence is also present in our approach; compare Eqs. (\ref{leftvertex5})
and (\ref{leftvertexs5}). Since the derivation of the vertices  that
enter the expression for the memory function for the self-intermediate 
scattering function is more straightforward, the presence of this correspondence 
serves as an additional check for the derivation presented above.

Combining the three steps and
taking the thermodynamic limit we arrive
at the following expression for the irreducible memory function,
\begin{eqnarray}\label{memfction}
M^{\mathrm{irr}}(q;t) &=&
\frac{\rho 
\omega_{\parallel}(q)}{2} \int \frac{d\mathbf{q}_1 d\mathbf{q}_2}{(2\pi)^3}
\delta(\mathbf{q}-\mathbf{q}_1-\mathbf{q}_2) \\ \nonumber && \times
\left(\hat{\mathbf{q}}\cdot\left[ \mathbf{q}_1 \mathcal{C}(q_1) + 
\mathbf{q}_2 \mathcal{C}(q_2)
\right]\right)^2 F(q_1;t)F(q_2;t).
\end{eqnarray}

Expression (\ref{memfction}) resembles the corresponding expression derived within 
the standard mode-coupling theory. It describes a dynamic feedback mechanism:
the time-delayed internal friction arising due to interparticle interactions
decays due to the relaxation of the two-particle density, which is included at
the level of factorization approximation (\ref{fac4to2}). Thus, slow decay of
density fluctuations feeds back into slow decay of the irreducible memory function.

Eq. (\ref{memfction}) incorporates the non-equilibrium nature of the active system
through the presence of the velocity correlations described through function 
$\omega_{\parallel}(q)$. This function sets the overall scale of the memory 
function. More importantly, it enters 
into the expression for new function $\mathcal{C}(q)$,
Eq. (\ref{newc}), and thus contributes to the vertices, which
weight relative contributions of density fluctuations to the memory function. 
As we will see in the next section, it is the presence of the velocity
correlations in the vertices that influences the location of the 
ergodicity breaking transition.

\section{Ergodicity breaking transition and long-time dynamics 
close to the transition}\label{sec:glassydyn}

In general, one needs to solve the combined set of Eqs. (\ref{memfunirr3}) and 
(\ref{memfction}) numerically. However, as we show in this section, even without
a numerical solution we can draw some general conclusions from these equations.

First, let us assume that the system, as described by Eqs. (\ref{memfunirr3}) and 
(\ref{memfction}), undergoes an ergodicity breaking transition. At such a 
transition, the intermediate scattering function does not decay to zero. The
long-time limit of this function, $\lim_{t\to\infty} F(q;t) = F(q) \equiv f(q) S(q)$,
is the order parameter of the non-ergodic state. Following the procedure used
in the standard mode-coupling theory \cite{Goetzebook} we can derive a self-consistent
equation for the normalized order parameter, $f(q)$, 
\begin{eqnarray}\label{fq}
\frac{f(q)}{1-f(q)} = m(q)
\end{eqnarray}
where $m(q)$ is given by the following equation
\begin{eqnarray}\label{mq}
m(q) &=& \frac{\rho}{2q^2} \int \frac{d\mathbf{q}_1 d\mathbf{q}_2}{(2\pi)^3}
\delta(\mathbf{q}-\mathbf{q}_1-\mathbf{q}_2) S(q) S(q_1) S(q_2) \nonumber \\ && \times
\left(\hat{\mathbf{q}}\cdot\left[ \mathbf{q}_1 \mathcal{C}(q_1) + 
\mathbf{q}_2 \mathcal{C}(q_2)
\right]\right)^2 f(q_1)f(q_2).
\end{eqnarray}

The self-consistent equation for the order parameter is very similar to the equation
derived in the standard mode-coupling theory. The only difference is that the role of 
the function describing the effective interaction, which in the standard mode-coupling
theory is played by the direct correlation function, is now played by the new function
$\mathcal{C}(q)$ which involves both the steady-state structure factor, $S(q)$, and
the function describing spatial correlations of the particles' velocities,
$\omega_{\parallel}(q)$. The presence of $\omega_{\parallel}(q)$ in the vertex
function means that the location of the ergodicity breaking transition is not
exclusively determined by the local steady-state structure. 

Finally, to look at the long-time dynamics in the ergodic phase but close to
the ergodicty breaking transition, it is convenient to introduce a normalized
correlator, $\phi(q;t) = F(q;t)/S(q)$. The equation of motion, Eq. (\ref{memfunirr3}),
in the Laplace space and re-written in terms of the correlator $\phi(q;z)$, has the 
following form,
\begin{eqnarray}\label{phiqz}
\frac{\phi(q;z)}{1-z\phi(q;z)} = 
\frac{z+\tau_p^{-1}+(\omega_{\parallel}(q)q^2/S(q)) m(q;z)}
{\omega_{\parallel}(q)q^2/S(q)},
\end{eqnarray}
where $m(q;z)$ is the Laplace transform of the reduced memory function $m(q;t)$,
\begin{eqnarray}\label{mqt}
m(q;t) &=& \frac{\rho}{2q^2} \int \frac{d\mathbf{q}_1 d\mathbf{q}_2}{(2\pi)^3}
\delta(\mathbf{q}-\mathbf{q}_1-\mathbf{q}_2) S(q) S(q_1) S(q_2) \nonumber \\ && \times
\left(\hat{\mathbf{q}}\cdot\left[ \mathbf{q}_1 \mathcal{C}(q_1) + 
\mathbf{q}_2 \mathcal{C}(q_2)
\right]\right)^2 \phi(q_1;t)\phi(q_2;t).
\end{eqnarray}
Near the ergodicity breaking transition, for small $z$ the memory function $m(q;z)$ 
becomes very large. Thus, for small $z$ (\textit{i.e.} for long times), we can
approximate Eq. (\ref{phiqz}) by the following one,
\begin{eqnarray}\label{phiqz2}
\frac{\phi(q;z)}{1-z\phi(q;z)} = m(q;z).
\end{eqnarray}
Eq. (\ref{phiqz2}) has the structure identical to that of the equation of motion for the 
long-time dynamics near the ergodicity breaking transition described by the standard 
mode-coupling theory. This means that all analytical results based of the
standard mode-coupling theory carry over to the present theory for the
dynamics of the active system. In particular, the only quantity that one needs to
do in order to predict the so-called mode-coupling exponents is to calculate the
exponent parameter $\lambda$ \cite{Goetzebook}. This parameter can be calculated
from the solution of the self-consistent equation for the order parameter,
Eqs. (\ref{fq}-\ref{mq}) at the ergodicity breaking transition.

The reduced memory function $m(q;z)$ differs from the corresponding quantity 
of the standard mode-coupling theory by the presence of the new function 
$\mathcal{C}(q)$, which involves $S(q)$ and $\omega_{\parallel}(q)$.  Again,
due to the fact that the correlation function of particles' velocities enters
into $m(q;t)$, the static structure factor does not completely determine the 
system's dynamics.

\section{Discussion}\label{sec:disc}

We presented here the details of the derivation of a recently proposed
theory for the dynamics of dense athermal active systems. The theory, in a
natural way, identifies a new function that influences both short- and long-time
dynamics. This new function quantifies spatial correlations of the particles' 
velocities in the stationary state. The presence of the new function implies 
that the dynamics is not determined by the local structure only. 

The influence of the additional steady-state function on the dynamics allows 
us to describe an un-expected result obtained in computer simulations \cite{activemct}.
We found that for a range of single-particle effective temperatures upon increasing
departure from equilibrium (by increasing the relaxation time of the self-propulsion),
the local structure becomes monotonically more pronounced whereas the long-time
dynamics first speeds up and then slows down. We showed in Ref. \cite{activemct} 
that our theory, combined with steady-state correlation functions obtained 
from computer simulations of a model single-component, moderately supercooled system, 
is able to describe qualitatively correctly the
non-monotonic dependence of the long-time dynamics on the persistence time
of the self-propulsion. 

Our theory shares the most important approximation, the factorization approximation,
with the mode-coupling theory of glassy dynamics. We showed that if there is an
ergodicity breaking transition, our theory predicts that the dynamics upon approaching 
this transition is qualitatively similar to that predicted by the standard mode-coupling
theory close to the corresponding mode-coupling transition. 

In future work we would like to extend the present approach to binary mixtures
in order to test the theory quantitatively against computer simulations, along the
lines of our earlier test of the standard mode-coupling theory \cite{bdmct}. 
Of particular interest would be to check whether the present theory can predict the 
location of the dynamic crossover in an active system more accurately than the standard 
mode-coupling theory can predict the location of the crossover in an equilibrium 
(thermal) system.

In addition, we would like to investigate the very fundamental approximation of
the theory concerning the absence of the steady-state currents. In this context, we
would like to develop and test a theory for effective temperatures in the active system. 

Finally, in real experimental active systems both self-propulsion and
thermal Brownian forces are very often present. For this reason, we would like to 
develop a more general theory that would be able to account for the presence of the 
Brownian noise.

\section*{Acknowledgments} 
The research in Montpellier was supported by funding
from the European Research Council under the European
Union's Seventh Framework Programme (FP7/2007-2013) / ERC Grant agreement No 306845.
I thank Ludovic Berthier for many
stimulating discussions and Elijah Flenner for comments on the
manuscript. 
I gratefully acknowledge the support of NSF Grant No.~CHE 121340.

\appendix
\section{Alternative  approximation for 
$\mathcal{Q}_{\text{lss}}\Omega\mathcal{Q}_{\text{lss}}$}\label{ap:alter}

We start by noting that due to Eqs. (\ref{right}) and (\ref{left}) the quantity 
we need to calculate is a specific matrix element of the inverse operator 
$\left(z-\mathcal{Q}_{\text{lss}}\Omega\mathcal{Q}_{\text{lss}}\right)^{-1}$
\begin{eqnarray}\label{matrixel}
&& \sum_i \boldsymbol{\nabla}_i \cdot
\int  d\mathbf{f}_1 ... d\mathbf{f}_N
\left(\mathbf{f}_i - \left<\mathbf{f}_i\right>_{\text{lss}}\right)
\left(z-\mathcal{Q}_{\text{lss}}\Omega\mathcal{Q}_{\text{lss}}\right)^{-1}
\nonumber \\ && \times 
\sum_j 
\left(\mathbf{f}_j - \left<\mathbf{f}_j\right>_{\text{lss}}\right)
P_N^{\text{ss}}(\mathbf{r}_1,\mathbf{f}_1,...,\mathbf{r}_N,\mathbf{f}_N)
\cdot\boldsymbol{\nabla}_j 
\nonumber \\ & \equiv &
\sum_{i,j} \boldsymbol{\nabla}_i \cdot
\left< \left(\mathbf{f}_i - \left<\mathbf{f}_i\right>_{\text{lss}}\right)
\left(z-\mathcal{Q}_{\text{lss}}\Omega\mathcal{Q}_{\text{lss}}\right)^{-1} 
\left(\mathbf{f}_j - \left<\mathbf{f}_j\right>_{\text{lss}}\right)\right>_{\text{lss}}
\nonumber \\ && 
\times 
P_N^{\text{ss}}(\mathbf{r}_1,...,\mathbf{r}_N)\cdot \boldsymbol{\nabla}_m. 
\end{eqnarray}

Following the spirit of the first Enskog approximation used in the kinetic theory
we introduce the following approximation,
\begin{widetext}
\begin{eqnarray}\label{matrixelapp}
&&
\sum_{i,j} \boldsymbol{\nabla}_i \cdot
\left< \left(\mathbf{f}_i - \left<\mathbf{f}_i\right>_{\text{lss}}\right)
\left(z-\mathcal{Q}_{\text{lss}}\Omega\mathcal{Q}_{\text{lss}}\right)^{-1} 
\left(\mathbf{f}_j - \left<\mathbf{f}_j\right>_{\text{lss}}\right)\right>_{\text{lss}}
P_N^{\text{ss}}(\mathbf{r}_1,...,\mathbf{r}_N)\cdot \boldsymbol{\nabla}_m
\nonumber \\ & \approx &
 \sum_{i,j,l,m} \boldsymbol{\nabla}_i
\cdot\left(\left<\mathbf{f}_i \mathbf{f}_j\right>_{\text{lss}} 
- \left<\mathbf{f}_i\right>_{\text{lss}} \left<\mathbf{f}_j\right>_{\text{lss}}\right)
\cdot
\left[ \left<\left(\mathbf{f}_j - \left<\mathbf{f}_j\right>_{\text{lss}}\right)
\left(z-\mathcal{Q}_{\text{lss}}\Omega\mathcal{Q}_{\text{lss}}\right)
\left(\mathbf{f}_l - \left<\mathbf{f}_l\right>_{\text{lss}}\right)\right>_{\text{lss}}
\right]^{-1}
\nonumber \\ && \cdot
\left(\left<\mathbf{f}_l \mathbf{f}_m\right>_{\text{lss}} 
- \left<\mathbf{f}_l\right>_{\text{lss}} \left<\mathbf{f}_m\right>_{\text{lss}}\right)
P_N^{\text{ss}}(\mathbf{r}_1,...,\mathbf{r}_N)
\cdot \boldsymbol{\nabla}_m 
\end{eqnarray}
Then we note that 
\begin{eqnarray}\label{matrixel2}
&& \left<\left(\mathbf{f}_j - \left<\mathbf{f}_j\right>_{\text{lss}}\right)
\mathcal{Q}_{\text{lss}}\Omega\mathcal{Q}_{\text{lss}}
\left(\mathbf{f}_l - \left<\mathbf{f}_l\right>_{\text{lss}}\right)\right>_{\text{lss}}
= 
\left<(\mathbf{f}_j - \left<\mathbf{f}_j\right>_{\text{lss}}) 
\Omega \mathbf{f}_l\right>_{\text{lss}}
\\ \nonumber &=&
\left[P_N^{\text{ss}}(\mathbf{r}_1,...,\mathbf{r}_N)\right]^{-1}
\int  d\mathbf{f}_1 ... d\mathbf{f}_N 
(\mathbf{f}_j - \left<\mathbf{f}_j\right>_{\text{lss}})
\Omega \mathbf{f}_l
P_N^{\text{ss}}(\mathbf{r}_1,\mathbf{f}_1,...,\mathbf{r}_N,\mathbf{f}_N)= 
\tau_p^{-1}
\left(\left<\mathbf{f}_j \mathbf{f}_l\right>_{\text{lss}} 
- \left<\mathbf{f}_j\right>_{\text{lss}} \left<\mathbf{f}_l\right>_{\text{lss}}\right)
- 2 D_f \delta_{jl} \delta_{\alpha\beta}. 
\end{eqnarray}

Eqs. (\ref{matrixelapp}-\ref{matrixel2}) lead to the following approximate effective
evolution operator,
\begin{eqnarray}\label{Omegaeffalt}
\Omega^{\text{eff}}(z) &=& \xi_0^{-2} \sum_{i,j,l,m} 
\boldsymbol{\nabla}_i \cdot
\left(\left<\mathbf{f}_i \mathbf{f}_j\right>_{\text{lss}} 
- \left<\mathbf{f}_i\right>_{\text{lss}} \left<\mathbf{f}_j\right>_{\text{lss}}\right)
\cdot\left[ (z-\tau_p^{-1}) \left(\left<\mathbf{f}_j \mathbf{f}_l\right>_{\text{lss}} 
- \left<\mathbf{f}_j\right>_{\text{lss}} \left<\mathbf{f}_l\right>_{\text{lss}}\right)
+ 2 D_f \delta_{jl} \delta_{\alpha\beta}
\right]^{-1}
\nonumber \\ && \cdot
\left(\left<\mathbf{f}_l \mathbf{f}_m\right>_{\text{lss}} 
- \left<\mathbf{f}_l\right>_{\text{lss}} \left<\mathbf{f}_m\right>_{\text{lss}}\right)
\cdot
\left[-\mathbf{F}_j^{\text{eff}}+\boldsymbol{\nabla}_j\right].
\end{eqnarray}
\end{widetext}

It can be showed that expression (\ref{Omegaeffalt}) reproduces correctly 
the third time derivative of the scattering functions at $t=0$. However, at
present it seems that additional approximations are necessary in order to 
use (\ref{Omegaeffalt}) as the starting point for the derivation of the memory
function representation for the scattering functions.

\section{Tagged particle (tracer) density fluctuations}\label{ap:self}

The derivation of the approximate equation of motion for the self-intermediate
scattering function, which describes the time dependence of the tagged particle
(tracer) density fluctuations follows the steps of the derivation presented 
in Sections \ref{sec:memf}-\ref{sec:mct}. In this Appendix we present all the 
formulas but, in the interest of brevity, we limit the comments to the minimum.

We start be rewriting the Laplace transform of the self-intermediate scattering function
in terms of the approximate evolution operator (\ref{Omegaeff}),
\begin{eqnarray}\label{Fsqz}
\lefteqn{\mathcal{LT}\left[F_s(q;t)\right](z) \equiv 
F_s(q;z)= } \nonumber \\ && 
\left<n_s(\mathbf{q})\left(z-\Omega\right)^{-1} n_s(-\mathbf{q})
\right> =  \nonumber \\ && \left<n_s(\mathbf{q}) 
\left(z-\Omega^{\text{eff}}(z)\right)^{-1} 
n_s(-\mathbf{q})\right>_\mathbf{r}.
\end{eqnarray}

To derive the memory function representation of $F(q;z)$ we use the projection
operator approach \cite{Goetzebook,CHess,SL}. We define a projection operator 
on the microscopic tagged particle density
\begin{equation}\label{Ps}
\mathcal{P}_s = ... \left. n_s(-\mathbf{q})\right>_\mathbf{r}
\left< n_s(\mathbf{q}) ... \right. .
\end{equation}
Note that $\left< n_s(\mathbf{q})n_s(-\mathbf{q})\right>_\mathbf{r}\equiv 1$ and
thus we do not need to include a normalization constant in definition (\ref{Ps}).

Next, we use the identity
\begin{eqnarray}\label{identitys}
\frac{1}{z-\Omega^{\text{eff}}(z)} 
&=& \frac{1}{z - \Omega^{\text{eff}}(z)\mathcal{Q}_s} 
\\ \nonumber && + 
\frac{1}{z - \Omega^{\text{eff}}(z)\mathcal{Q}_s}\Omega^{\text{eff}}(z)\mathcal{P}_s
\frac{1}{z-\Omega^{\text{eff}}(z)},
\end{eqnarray} 
where $\mathcal{Q}_s$ is the projection on the space orthogonal to that 
spanned by the microscopic tagged particle density, 
\begin{equation}
\mathcal{Q}_s = \mathcal{I} - \mathcal{P}_s,
\end{equation}
to rewrite the Laplace transform of the time derivative of $F_s(q;t)$ in the following 
way
\begin{widetext}
\begin{eqnarray}\label{timeders}
&& \mathcal{LT}[\partial_t F_s(q;t)](z)= 
\left<n_s(\mathbf{q}) \Omega^{\text{eff}}(z) 
\frac{1}{z-\Omega^{\text{eff}}(z)} n_s(-\mathbf{q}) \right>_\mathbf{r} 
= \left<n_s(\mathbf{q}) \Omega^{\text{eff}}(z) \mathcal{P}_s 
\frac{1}{z-\Omega^{\text{eff}}(z)} n_s(-\mathbf{q}) \right>_\mathbf{r}
\nonumber \\ && + \left<n_s(\mathbf{q}) \Omega^{\text{eff}}(z) \mathcal{Q}_s 
\frac{1}{z-\Omega^{\text{eff}}(z)} n_s(-\mathbf{q}) \right>_\mathbf{r} 
=
\left<n_s(\mathbf{q}) \Omega^{\text{eff}}(z) n_s(-\mathbf{q})\right>_\mathbf{r}
\left< n_s(\mathbf{q}) \frac{1}{z-\Omega^{\text{eff}}(z)} n_s(-\mathbf{q}) 
\right>_\mathbf{r} 
\nonumber \\ && + 
\left<n_s(\mathbf{q}) \Omega^{\text{eff}}(z) \mathcal{Q}_s 
\frac{1}{z - \mathcal{Q}_s\Omega^{\text{eff}}(z)\mathcal{Q}_s} 
\mathcal{Q}_s \Omega^{\text{eff}}(z) n_S(-\mathbf{q})\right>_\mathbf{r}
\left< n_s(\mathbf{q})\frac{1}{z-\Omega^{\text{eff}}(z)}
n_s(-\mathbf{q}) \right>_\mathbf{r}.
\end{eqnarray}
\end{widetext}
The tagged particle frequency matrix, $\mathcal{H}_s(q;z)$, is defined in terms
of the matrix element of the effective evolution operator, 
\begin{eqnarray}\label{freqmats1}
\left<n_s(\mathbf{q}) \Omega^{\text{eff}}(z) n_s(-\mathbf{q})\right>_\mathbf{r}
= -q^2 \mathcal{H}_s(q;z),
\end{eqnarray}
where
\begin{eqnarray}\label{freqmats2}
\mathcal{H}_s(q;z) = 
\frac{ \hat{\mathbf{q}}\cdot
\left< \left<\mathbf{f}_1 \mathbf{f}_1\right>_{\text{lss}} 
- \left<\mathbf{f}_1\right>_{\text{lss}} \left<\mathbf{f}_1\right>_{\text{lss}}
\right>_\mathbf{r} 
\cdot\hat{\mathbf{q}}}
{\xi_0^2\left(z+\tau_p^{-1}\right)} = \frac{\omega_{\parallel}(\infty)}{z+\tau_p^{-1}}.
\nonumber \\
\end{eqnarray}
The tagged particle memory matrix is defined as follows,
\begin{eqnarray}\label{memfunreds1}
&& \left<n_s(\mathbf{q}) \Omega^{\text{eff}}(z) \mathcal{Q}_s 
\frac{1}{z - \mathcal{Q}_s\Omega^{\text{eff}}(z)\mathcal{Q}_s} 
\mathcal{Q}_s \Omega^{\text{eff}}(z) n_s(-\mathbf{q})\right>_{\mathbf{r}} =
\nonumber \\ && q^2 \mathcal{M}_s(q;z).
\end{eqnarray}
Explicitly, 
the memory matrix is given by the following expression
\begin{widetext}
\begin{eqnarray}\label{memfunreds2}
&&  \mathcal{M}_s(q;z) = 
\left(\xi_0^4\left(z+\tau_p^{-1}\right)^2\right)^{-1}
\hat{\mathbf{q}}\cdot\left< \sum_{j} e^{-i\mathbf{q}\cdot\mathbf{r}_1}
\left(\left<\mathbf{f}_1 \mathbf{f}_j\right>_{\text{lss}} 
- \left<\mathbf{f}_1\right>_{\text{lss}} 
\left<\mathbf{f}_j\right>_{\text{lss}}\right)\cdot
\right. \nonumber \\ && \times \left.
\left[-\boldsymbol{\nabla}_j + \mathbf{F}^{\text{eff}}_j\right]
\mathcal{Q}_s
\frac{1}{z - \mathcal{Q}_s\Omega^{\text{eff}}(z)\mathcal{Q}_s}
\mathcal{Q}_s \sum_{l} \boldsymbol{\nabla}_l \cdot
\left(\left<\mathbf{f}_l \mathbf{f}_1\right>_{\text{lss}} 
- \left<\mathbf{f}_l\right>_{\text{lss}} \left<\mathbf{f}_1\right>_{\text{lss}}\right)
e^{i\mathbf{q}\cdot\mathbf{r}_1} \right>_{\mathbf{r}}\cdot\hat{\mathbf{q}}
\end{eqnarray}
\end{widetext}
To define the irreducible tagged particle memory matrix we start by introducing
an irreducible evolution operator for the tagged particle motion,
\begin{equation}\label{Omegairrs1}
\Omega^{\text{irr}}_s(z) = \mathcal{Q}_s\Omega^{\text{eff}}(z)\mathcal{Q}_s -
\delta \Omega^{\text{eff}}_s(z)
\end{equation}
where the subtraction term  $\delta \Omega^{\text{eff}}_s(z)$ reads 
\begin{widetext}
\begin{eqnarray}\label{Omegairrs2}
\delta \Omega^{\text{irr}}_s(z) &=& \left. \mathcal{Q}_s
\left(\xi_0^4\left(z+\tau_p^{-1}\right)^2\right)^{-1}\sum_{i}
\boldsymbol{\nabla}_i \cdot \left(\left<\mathbf{f}_i \mathbf{f}_1\right>_{\text{lss}} 
- \left<\mathbf{f}_i\right>_{\text{lss}} \left<\mathbf{f}_1\right>_{\text{lss}}\right)
e^{i\mathbf{q}\cdot \mathbf{r}_1}\right>_{\mathbf{r}}\cdot\hat{\mathbf{q}}
\left(\mathcal{H}_s(q;z)\right)^{-1}
\nonumber \\ && \hat{\mathbf{q}}\cdot
\left< \sum_{l} e^{-i\mathbf{q}\cdot \mathbf{r}_1}
\left(\left<\mathbf{f}_1 \mathbf{f}_l\right>_{\text{lss}} 
- \left<\mathbf{f}_1\right>_{\text{lss}} \left<\mathbf{f}_l\right>_{\text{lss}}\right)
\cdot \left(\boldsymbol{\nabla}_l-\beta\mathbf{F}^{\text{eff}}_l\right)\mathcal{Q}_s 
\right.
\end{eqnarray}
Next, we define the tagged particle irreducible memory matrix 
$\mathcal{M}^{\text{irr}}_s(q;z)$, which is given by the expression analogous to 
Eq. (\ref{memfunreds2}) but with the projected evolution operator
$\mathcal{Q}_s\Omega^{\text{eff}}(z)\mathcal{Q}_s$ replaced by 
irreducible evolution  operator $\Omega^{\text{irr}}_s(z)$,
\begin{eqnarray}\label{memfunirrs1}
&&  \mathcal{M}^{\text{irr}}_s(q;z) = 
\left(\xi_0^4\left(z+\tau_p^{-1}\right)^2\right)^{-1}
\hat{\mathbf{q}}\cdot\left< \sum_{j} e^{-i\mathbf{q}\cdot\mathbf{r}_1}
\left(\left<\mathbf{f}_1 \mathbf{f}_j\right>_{\text{lss}} 
- \left<\mathbf{f}_1\right>_{\text{lss}} 
\left<\mathbf{f}_j\right>_{\text{lss}}\right)\cdot
\right. \nonumber \\ && \times \left.
\left[-\boldsymbol{\nabla}_j + \mathbf{F}^{\text{eff}}_j\right]
\mathcal{Q}_s
\frac{1}{z - \Omega^{\text{irr}}_s(z)}
\mathcal{Q}_s \sum_{l} \boldsymbol{\nabla}_l \cdot
\left(\left<\mathbf{f}_l \mathbf{f}_1\right>_{\text{lss}} 
- \left<\mathbf{f}_l\right>_{\text{lss}} \left<\mathbf{f}_1\right>_{\text{lss}}\right)
e^{i\mathbf{q}\cdot\mathbf{r}_1} \right>_{\mathbf{r}}\cdot\hat{\mathbf{q}}.
\end{eqnarray}
\end{widetext}
Finally, we use an identity similar to Eq. (\ref{identitys}),
\begin{eqnarray}\label{identitys2}
\lefteqn{\frac{1}{z-\mathcal{Q}_s\Omega^{\text{eff}}(z)\mathcal{Q}_s} 
= \frac{1}{z - \Omega^{\text{irr}}_s(z)} }
\\ \nonumber && + 
\frac{1}{z - \Omega^{\text{irr}}_s(z)}\delta\Omega^{\text{eff}}_s(z) 
\frac{1}{z-\mathcal{Q}_s\Omega^{\text{eff}}(z)\mathcal{Q}_s},
\end{eqnarray} 
and we derive a relation between $\mathcal{M}_s(q;z)$ and 
$\mathcal{M}^{\text{irr}}_s(q;z)$,
\begin{equation}\label{redirrs}
\mathcal{M}_s(q;z)=\mathcal{M}^{\text{irr}}_s(q;z)-\mathcal{M}^{\text{irr}}_s(q;z)
\mathcal{H}^{-1}_s(q;z)\mathcal{M}_s(q;z).
\end{equation}
The memory function representation for the self-intermediate scattering functions reads,
\begin{eqnarray}\label{fqzmfirrs}
F_s(q;z) = \frac{1}{z+\frac{q^2\mathcal{H}_s(q;z)}
{1+\mathcal{M}^{\text{irr}}_s(q;z)/\mathcal{H}_s(q;z)}}.
\end{eqnarray}
Again, we introduce the tagged particle (self) memory function,
\begin{eqnarray}\label{memfunirrs2}
M^{\text{irr}}_s(q;z) = 
\left(z+\tau_p^{-1}\right)^2\mathcal{M}^{\text{irr}}(q;z)_s/\omega_{\parallel}(\infty),
\end{eqnarray}
we rewrite Eq. (\ref{fqzmfirrs}) in the time domain,
\begin{eqnarray}\label{memfunirrs3}
\lefteqn{
\partial^2_t F_s(q;t) +\tau_p^{-1} \partial_t  F_s(q;t) =  }
\\ \nonumber &&  
-\omega_{\parallel}(\infty) q^2 F_s(q;t)
- \int_0^t dt' M^{\text{irr}}_s(q;t-t') \partial_{t'} F_s(q;t'),
\end{eqnarray}
where $M^{\text{irr}}_s(q;t)$ is the inverse Laplace transform of
$M^{\text{irr}}_s(q;z)$. We note that for evolution on the time scale longer than the 
persistence time the second time derivative term in Eq. (\ref{memfunirrs3}) can be 
neglected. The resulting equation becomes equivalent to that describing the 
self-intermediate scattering function of a thermal Brownian system 
with a short time 
self-diffusion coefficient $\omega_{\parallel}(\infty)\tau_p$. The last observation
suggests that, in order to quantify the effect of the time-delayed friction
(represented by the memory function term) on the evolution of the active system,
one should normalize the time scale by the term describing the short-time
dynamics, $\omega_{\parallel}(\infty)\tau_p$.

Finally, we note that Eq. (\ref{memfunirrs3}) leads to the following
expression for the second time derivative of the self-intermediate scattering
function at $t=0$, 
\begin{equation}\label{tderFsqt}
\left. \partial^2_t F_s(q;t)\right|_{t=0} = -\omega_{\parallel}(\infty) q^2 F_s(q;t=0) 
\equiv -\omega_{\parallel}(\infty) q^2,
\end{equation}
which agrees with the exact result (\ref{Fsqtst1}).

To derive an approximate expression for  tagged particle memory function,
we again follow the three steps \cite{SL}, with some modifications due to the
difference between the tracer (particle number 1) and all other particles.

First, we project onto the subspace of joint tagged and fluid particle density,
\begin{widetext}
\begin{eqnarray}\label{projdenjoint}
\lefteqn{\hat{\mathbf{q}}\cdot\left< \sum_{j} e^{-i\mathbf{q}\cdot\mathbf{r}_1}
\left(\left<\mathbf{f}_1 \mathbf{f}_j\right>_{\text{lss}} 
- \left<\mathbf{f}_1\right>_{\text{lss}} 
\left<\mathbf{f}_j\right>_{\text{lss}}\right)\cdot
\left[-\boldsymbol{\nabla}_j + \mathbf{F}^{\text{eff}}_j\right]
\mathcal{Q}_s \right. }
\nonumber \\ \nonumber && \approx \sum_{\mathbf{q}_1,...,\mathbf{q}_4}
\hat{\mathbf{q}}\cdot\left< \sum_{j} e^{-i\mathbf{q}\cdot\mathbf{r}_1}
\left(\left<\mathbf{f}_1 \mathbf{f}_j\right>_{\text{lss}} 
- \left<\mathbf{f}_1\right>_{\text{lss}} 
\left<\mathbf{f}_j\right>_{\text{lss}}\right)\cdot
\left[-\boldsymbol{\nabla}_j + \mathbf{F}^{\text{eff}}_j\right]
\mathcal{Q}_s n_{s2}(-\mathbf{q}_1,-\mathbf{q}_2)\right>_{\mathbf{r}} 
\\ && \times 
\left[\left<\mathcal{Q}_s n_{s2}(\mathbf{q}_1,\mathbf{q}_2) \mathcal{Q}_s 
n_{s2}(-\mathbf{q}_3,-\mathbf{q}_4)
\right>_{\mathbf{r}}\right]^{-1}\left< \mathcal{Q}_1 n_{s2}(\mathbf{q}_3,\mathbf{q}_4)
\right.
\end{eqnarray}
\end{widetext}
Here $n_{s2}(\mathbf{q}_1,\mathbf{q}_2)$ is the Fourier transform of the microscopic 
joint tagged and fluid particle density,
\begin{equation}\label{ns2def}
n_{s2}(\mathbf{q}_1,\mathbf{q}_2) = \sum_{m>2} 
e^{-i\mathbf{q}_1\cdot\mathbf{r}_1-i\mathbf{q}_2\cdot\mathbf{r}_m},
\end{equation}
and $\left[\left<\mathcal{Q}_s n_{s2}(\mathbf{q}_1,\mathbf{q}_2) \mathcal{Q}_s 
n_{s2}(-\mathbf{q}_3,-\mathbf{q}_4)
\right>_{\mathbf{r}}\right]^{-1}$ is the inverse of the correlation matrix of 
microscopic joint densities. 
It should be noted that in Eq. (\ref{projdenjoint}) 
we are using parts of the microscopic joint density that are orthogonal to the 
microscopic tagged particle density, $\mathcal{Q}_1 n_{s2}(\mathbf{q}_1,\mathbf{q}_2)$. 
The presence
of the operator $\mathcal{Q}_s$ is necessary for the existence of the inverse
of the correlation matrix 
$\left<\mathcal{Q}_s n_{s2}(\mathbf{q}_1,\mathbf{q}_2) \mathcal{Q}_s 
n_{s2}(-\mathbf{q}_3,-\mathbf{q}_4)\right>_{\mathbf{r}}$ \cite{Anderson}. 

Second, projecting onto joint tagged-fluid particle densities leads to an expression
involving a four-particle correlation function which is factorized. This 
factorization approximation is the main approximation involved in the present
derivation, 
\begin{widetext}
\begin{eqnarray}\label{facs4tos2}
\lefteqn{\mathcal{LT}^{-1}\left[\left< \mathcal{Q}_s n_{s2}(\mathbf{q}_1,\mathbf{q}_2) 
\left(z-\Omega^{\text{irr}}_s(z)\right)^{-1} \mathcal{Q}_s 
n_{s2}(-\mathbf{q}_3,-\mathbf{q}_4)\right>_{\mathbf{r}}\right] \approx}
\\ \nonumber && 
\mathcal{LT}^{-1}\left[\left< n_s(\mathbf{q}_1) 
\left(z-\Omega^{\text{eff}}(z)\right)^{-1} n_s(-\mathbf{q}_3)\right>_{\mathbf{r}}\right]
\mathcal{LT}^{-1}\left[\left< n(\mathbf{q}_2) 
\left(z-\Omega^{\text{eff}}(z)\right)^{-1} n(-\mathbf{q}_4)\right>_{\mathbf{r}}\right].
\end{eqnarray}
\end{widetext}
Consistently with
Eq. (\ref{facs4tos2}) we also factorize the steady-state correlation matrix
of microscopic joint densities and for its inverse we get
\begin{eqnarray}\label{facs4tos2ss}
&& \left[\left<\mathcal{Q}_s n_{s2}(\mathbf{q}_1,\mathbf{q}_2) \mathcal{Q}_s 
n_{s2}(-\mathbf{q}_3,-\mathbf{q}_4)
\right>_{\mathbf{r}}\right]^{-1} \approx
\\ \nonumber && 
\delta_{\mathbf{q}_1,\mathbf{q}_3}
\left<n(\mathbf{q}_2) n(-\mathbf{q}_4)\right>_{\mathbf{r}}^{-1}.
\end{eqnarray}

The final, third step, is concerned with the vertices. The left vertex reads,
\begin{eqnarray}\label{leftvertexdefs}
\lefteqn{ \mathcal{V}_{sl}(\mathbf{q};\mathbf{q}_1,\mathbf{q}_2) = }
\nonumber \\ && 
\xi_0^{-2} \hat{\mathbf{q}}\cdot\left< \sum_{j} e^{-i\mathbf{q}\cdot\mathbf{r}_1}
\left(\left<\mathbf{f}_1 \mathbf{f}_j\right>_{\text{lss}} 
- \left<\mathbf{f}_1\right>_{\text{lss}} \left<\mathbf{f}_j\right>_{\text{lss}}\right)
\cdot \right.
\nonumber \\ && \times \left.
\left[-\nabla_j + \mathbf{F}^{\text{eff}}_j\right]
\mathcal{Q}_s n_{s2}(-\mathbf{q}_1,-\mathbf{q}_2)\right>_{\mathbf{r}}
\nonumber \\ &=&
\xi_0^{-2} \hat{\mathbf{q}}\cdot\left< \sum_{j} e^{-i\mathbf{q}\cdot\mathbf{r}_i}
\left(\mathbf{f}_1 + \mathbf{F}_1\right)
\left(\mathbf{f}_j + \mathbf{F}_j\right)\cdot \right.
\nonumber \\ && \times \left.
\left[-\nabla_j + \mathbf{F}^{\text{eff}}_j\right]
\mathcal{Q}_s n_{s2}(-\mathbf{q}_1,-\mathbf{q}_2)\right>
\end{eqnarray}
Due to the presence of the projection operator $\mathcal{Q}_n$, 
the left vertex consists of two terms,
\begin{eqnarray}\label{leftvertexs1}
\lefteqn{ \mathcal{V}_{sl}(\mathbf{q};\mathbf{q}_1,\mathbf{q}_2) = }
\\ \nonumber 
&-& \xi_0^{-2} \hat{\mathbf{q}}\cdot\left< \sum_{j} e^{-i\mathbf{q}\cdot\mathbf{r}_1}
\left(\mathbf{f}_1 + \mathbf{F}_1\right)
\left(\mathbf{f}_j + \mathbf{F}_j\right)\cdot \right.
\\ \nonumber && \times \left.
\left[\nabla_j n_{s2}(-\mathbf{q}_1,-\mathbf{q}_2)\right]\right>
\\ \nonumber 
&+& \xi_0^{-2} \hat{\mathbf{q}}\cdot\left< \sum_{j} e^{-i\mathbf{q}\cdot\mathbf{r}_1}
\left(\mathbf{f}_1 + \mathbf{F}_1\right)
\left(\mathbf{f}_j + \mathbf{F}_j\right)\cdot \right.
\\ \nonumber && \times \left.
\left[\nabla_j n_s(-\mathbf{q})\right]\right>
\left<n_s(\mathbf{q})n_{s2}(-\mathbf{q}_1,-\mathbf{q}_2)\right>.
\end{eqnarray}
The second term can be expressed in terms of $\omega_{\parallel}(\infty)$ and 
the steady-state structure factor,
\begin{eqnarray}\label{leftvertexs2}
&&\xi_0^{-2} \hat{\mathbf{q}}\cdot\left< \sum_{j} e^{-i\mathbf{q}\cdot\mathbf{r}_1}
\left(\mathbf{f}_1 + \mathbf{F}_1\right)
\left(\mathbf{f}_j + \mathbf{F}_j\right)\cdot \right.
\nonumber \\ \nonumber && \times \left.
\left[\nabla_j n_s(-\mathbf{q})\right]\right>
\left<n(\mathbf{q})n_{s2}(-\mathbf{q}_1,-\mathbf{q}_2)\right>
\\ \nonumber & = &
i \omega_{\parallel}(\infty) \hat{\mathbf{q}}\cdot \left(\mathbf{q}_1+\mathbf{q}_2\right)
(S(q_2)-1)\delta_{\mathbf{q},\mathbf{q}_1+\mathbf{q}_2}.
\end{eqnarray}
The first term at the right-hand-side of Eq. (\ref{leftvertexs1}) can be
rewritten as follows,
\begin{eqnarray}\label{leftvertexs3}
&&
-\xi_0^{-2} \hat{\mathbf{q}}\cdot\left< \sum_{j} e^{-i\mathbf{q}\cdot\mathbf{r}_1}
\left(\mathbf{f}_1 + \mathbf{F}_1\right)
\left(\mathbf{f}_j + \mathbf{F}_j\right)\cdot \right.
\\ \nonumber && \times \left.
\left[\nabla_j n_{s2}(-\mathbf{q}_1,-\mathbf{q}_2)\right]\right>
\\ \nonumber & = & 
-\xi_0^{-2} \hat{\mathbf{q}}\cdot\left< \sum_{j} e^{-i\mathbf{q}\cdot\mathbf{r}_1}
\left(\mathbf{f}_1 + \mathbf{F}_1\right)
\left(\mathbf{f}_j + \mathbf{F}_j\right)\cdot \right.
\\ \nonumber && \times \left. 
\left[i\mathbf{q}_1 e^{i\mathbf{q}_1\cdot\mathbf{r}_1}\delta_{j1}\sum_{l>1}  
e^{i\mathbf{q}_2\cdot\mathbf{r}_l} + 
i\mathbf{q}_2 e^{i\mathbf{q}_2\cdot\mathbf{r}_j}  
e^{i\mathbf{q}_1\cdot\mathbf{r}_1} (1-\delta_{j1})\right]\right>
\\ \nonumber & = & 
-\xi_0^{-2} \hat{\mathbf{q}}\cdot\left< \sum_{j} e^{-i\mathbf{q}\cdot\mathbf{r}_1}
\left(\mathbf{f}_1 + \mathbf{F}_1\right)
\left(\mathbf{f}_j + \mathbf{F}_j\right)\cdot \right.
\\ \nonumber && \times \left. 
\left[i\mathbf{q}_1 e^{i\mathbf{q}_1\cdot\mathbf{r}_1}\delta_{j1}\sum_{l}  
e^{i\mathbf{q}_2\cdot\mathbf{r}_l} 
- i(\mathbf{q}_1+\mathbf{q}_2)\delta_{j1}e^{i\mathbf{q}_2\cdot\mathbf{r}_1}
\right. \right. \\ \nonumber && \left.\left. 
+ i\mathbf{q}_2 e^{i\mathbf{q}_2\cdot\mathbf{r}_j}  
e^{i\mathbf{q}_1\cdot\mathbf{r}_1}\right]\right>
\\ \nonumber & \approx & -i\left[
\omega_{\parallel}(\infty) S(q_2) \hat{\mathbf{q}}\cdot\mathbf{q}_1
- \omega_{\parallel}(\infty) 
\hat{\mathbf{q}}\cdot \left(\mathbf{q}_1+\mathbf{q}_2\right)
\right. 
\\ \nonumber && \left.
+ \omega_{\parallel}(q_2) \hat{\mathbf{q}}\cdot\mathbf{q}_2 
\right]
\delta_{\mathbf{q},\mathbf{q}_1+\mathbf{q}_2},
\end{eqnarray}
where in the last step the following approximation was used
\begin{eqnarray}\label{leftvertexs4}
&& \left< e^{-i\mathbf{q}\cdot\mathbf{r}_1}
\left(\mathbf{f}_1 + \mathbf{F}_1\right)
\left(\mathbf{f}_1 + \mathbf{F}_1\right) 
e^{i\mathbf{q}_1\cdot\mathbf{r}_1}\sum_l e^{i\mathbf{q}_2\cdot\mathbf{r}_l}\right>
\\ \nonumber && \approx  
\left< \left(\mathbf{f}_1 + \mathbf{F}_1\right)
\left(\mathbf{f}_1 + \mathbf{F}_1\right) \right> 
\left< e^{-i(\mathbf{q}-\mathbf{q}_1)\cdot\mathbf{r}_1} 
\sum_l e^{i\mathbf{q}_2\cdot\mathbf{r}_l}\right>
\\ \nonumber && = \omega_{\parallel}(\infty)S(q_2)
\end{eqnarray}
Combining Eqs. (\ref{leftvertexs2}) and (\ref{leftvertexs3}) we get the 
following approximate expression for the left vertex,
\begin{eqnarray}\label{leftvertexs5}
\lefteqn{ \mathcal{V}_{sl}(\mathbf{q};\mathbf{q}_1,\mathbf{q}_2) 
\approx -i S(q_2) \omega_{\parallel}(\infty)}
\nonumber \\ & \times & 
\hat{\mathbf{q}}\cdot\mathbf{q}_2 \left(
\frac{\omega_{\parallel}(q_2)}{\omega_{\parallel}(\infty)S(q_2)} - 1\right)
\nonumber \\ & \equiv & 
i \rho S(q_2) \omega_{\parallel}(q)
\hat{\mathbf{q}}\cdot\mathbf{q}_2 \mathcal{C}(q_2),
\end{eqnarray}
where function $ \mathcal{C}(q)=(1-1/S(q))/\rho$, Eq. (\ref{newc}) of the main text.
The right vertex can be analyzed in the same way. 

Combining the three steps and
taking the thermodynamic limit we arrive
at the following expression for the irreducible memory function for
the tagged particle motion
\begin{eqnarray}\label{memfctions}
M^{\mathrm{irr}}_s(q;t) &=&
\rho 
\omega_{\parallel}(\infty) \int \frac{d\mathbf{q}_1 d\mathbf{q}_2}{(2\pi)^3}
\delta(\mathbf{q}-\mathbf{q}_1-\mathbf{q}_2) \nonumber \\ & \times &
\left(\hat{\mathbf{q}}\cdot\mathbf{q}_2 \mathcal{C}(q_2)\right)^2 F_s(q_1;t)F(q_2;t).
\end{eqnarray}

As we mentioned in the main text of the article, the relation between the 
approximate expressions for the irreducible memory functions for the 
collective and tagged particle motion, Eqs. (\ref{memfction}) and (\ref{memfctions}), 
is the same as in the standard mode-coupling theory.


\begin{thebibliography}{99}
\bibitem{commentMCT} Mode-coupling theory uses, effectively, a mean-field
approach and thus its microscopic character may be questioned. However, it
is one of the very few approaches that can make detailed predictions for
time-dependent correlation functions using as the input the static (equilibrium
or steady-state) correlations.
\bibitem{commentSP} The motion of real swimming bacteria or self-propelled Janus
particles is force free but we follow previous studies \cite{tenHagen} and describe the 
self-propulsion as an effective internal driving force).
\bibitem{commentlong} 
We note that the subtraction term $\delta \Omega^{\text{irr}}(z)$ includes
only longitudinal components (proportional to $\hat{\mathbf{q}}$). This 
follows from the fact that reducible memory matrix $\mathcal{M}(q;z)$ is
longitudinal.
\end{thebibliography}
\end{document}